\documentclass[prx,aps,twocolumn,floatfix,showpacs,superscriptaddress]{revtex4-1}
\usepackage{amssymb}
\usepackage{amsmath}
\usepackage{graphicx}
\usepackage[colorlinks=true,linkcolor=blue,anchorcolor=red,citecolor=blue,urlcolor=blue]{hyperref}
\usepackage{pdfpages}

\begin{document}

\title{Weak antilocalization and localization in disordered and interacting Weyl semimetals}

\author{Hai-Zhou Lu}
\affiliation{Department of Physics, South University of Science and Technology of China, Shenzhen, China}
\affiliation{Department of Physics, The University of Hong Kong, Pokfulam Road,
Hong Kong, China}

\author{Shun-Qing Shen}
\affiliation{Department of Physics, The University of Hong Kong, Pokfulam Road,
Hong Kong, China}

\date{\today }

\begin{abstract}
Using the Feynman diagram techniques, we derive the finite-temperature conductivity and magnetoconductivity formulas from the quantum interference and electron-electron interaction, for a three-dimensional disordered Weyl semimetal. For a single valley of Weyl fermions, we find that the magnetoconductivity is negative and proportional to the square root of magnetic field at low temperatures, as a result of the weak antilocalization. By including the contributions from the weak antilocalization, Berry curvature correction, and Lorentz force, we compare the calculated magnetoconductivity with a recent experiment. The weak antilocalization always dominates the magnetoconductivity near zero field, thus gives one of the transport signatures for Weyl semimetals. In the presence of strong intervalley scattering and correlations, we expect a crossover from the weak antilocalization to weak localization. In addition, we find that the interplay of electron-electron interaction and disorder scattering always dominates the conductivity at low temperatures and leads to a tendency to localization. Finally, we present a systematic comparison of the transport properties of single-valley Weyl fermions, 2D massless Dirac fermions, and 3D conventional electrons.
\end{abstract}


\maketitle


\section{Introduction}

Weyl semimetal is a three-dimensional (3D) topological
state of matter, in which the conduction and valence energy bands
touch at a finite number of nodes \cite{Balents11physics}. The nodes
always appear in pairs, in each pair the quasiparticles (dubbed Weyl
fermions) carry opposite chirality and linear dispersion, much like
a 3D analog of graphene. The neutrino used to be a potential candidate
for the Weyl fermion, until its tiny mass was revealed. In the past few
years, a number of condensed matter systems have been suggested to
host Weyl fermions \cite{Wan11prb,Yang11prb,Burkov11prl,Delplace12epl,Jiang12pra,Young12prl,Xu11prl,Wang12prb,Singh12prb,Wang13prb,LiuJP14prb,Bulmash14prb}.
Most recently, the signatures of Weyl nodes have been observed by
angle-resolved photoemission spectroscopy, scanning tunneling microscopy, and time-domain terahertz spectroscopy
in (Bi$_{1-x}$In$_x$)$_2$Se$_3$ \cite{Brahlek12prl,Wu13natphys}, Na$_{3}$Bi \cite{Liu14sci,Xu15sci}, Cd$_{3}$As$_{2}$  \cite{Liu14natmat,Neupane14nc,Yi14srep,Borisenko14prl,Jeon14natmat}, and TlBiSSe \cite{Novak15prbr} (Strictly speaking, they are Dirac semimetals in which the paired Weyl nodes are degenerate \cite{Hosur12prl,Young12prl}).

Excellent electronic transport is anticipated in Weyl semimetals.
The Weyl nodes remain gapless unless being annihilated in pairs.
It is known that disorder may induce a semimetal to metal transition \cite{Fradkin86prb,Shindou10njp,Goswami11prl,Syzranov15prl}.
Nevertheless, metals may also exhibit ``insulating'' behaviors as a result of
disorder and quantum interference, i.e., Anderson localization \cite{Lee85rmp}.
In contrast, because of the symplectic symmetry \cite{HLN80ptp,Suzuura02prl} near each Weyl node, the Weyl fermions are immune from Anderson localization
and tend to be ``antilocalized'', in the absence of interaction and intervalley scattering. One of the signatures of the weak antilocalization is a negative magnetoconductivity, and has been observed recently in Bi$_{0.97}$Sb$_{0.03}$
\cite{Kim13prl} with a theoretical description based on a corrected semiclassical Boltzmann equation \cite{Kim14prb}, ZrTe$_5$ \cite{Li14arXiv}, and TaAs \cite{Huang15nc,ZhangCL15arXiv}. However, to include the weak (anti-)localization corrections, higher-order Feynman diagrams \cite{McCann06prl,Altshuler80prl,Fukuyama80jpsj,Lee85rmp} beyond the semiclassical transport theory have to be taken into account. A full three-dimensional calculation beyond the semiclassical \cite{Hosur12prl,Biswas14prb,Gorbar14prb,Kim14prb} and quasi-two-dimensional \cite{Garate12prb} regimes is still lacking for this paradigmatic system, in particular in the presence of many-body interaction and multi-valley effects.

In this work, we systematically
study the temperature and magnetic field dependences of the conductivity
of a two-valley Weyl semimetal. With the help of Feynman diagram techniques, we take into account high-order corrections
from the quantum interference as well as the interplay of interaction
and disorder beyond the semiclassical transport theory. We find that the low-temperature magnetoconductivity is negative and follows a square-root law in weak magnetic fields $B$, (i.e., a $-\sqrt{B}$ magnetoconductivity) (see Fig. \ref{fig:MC})
arising from the weak antilocalization, which is in consistence with the
experiments \cite{Kim13prl,Li14arXiv,Huang15nc,ZhangCL15arXiv}. However,
despite this magnetoconductivity signature of the weak antilocalization,
the temperature dependence of the conductivity still shows a tendency
to localization below a critical temperature,
as a result of weak many-body interaction (see Fig. \ref{fig:sigma-T}).
Moreover, intervalley scattering and correlation may also strengthen
the localization tendency (see Fig. \ref{fig:MC}). This work brings the
transport theory to the level of relevant experiments to detect signatures
of Weyl fermions in solid-state systems.

The paper is organized as follows. In Sec. \ref{sec:model}, we introduce the model that describes a two-valley Weyl semimetal in the presence of electron-electron interaction and disorder. Then we briefly present the Feynman diagrams for the conductivity. In Sec. \ref{sec:sigma-temperature}, we show the temperature dependence of the conductivity at low temperatures. We focus on the competition between the weak antilocalization due to the quantum interference and the localization arising from the interplay of interaction and disorder scattering. In Sec. \ref{sec:MC}, we present the $-\sqrt{B}$ magnetoconductivity from the weak antilocalization of a single valley of Weyl fermions. Then we discuss the crossover to the weak localization as a result of the intervalley scattering and correlation. We also compare with a recent experiment, by including the magnetoconductivity contributions from the weak antilocalization, Berry curvature correction, and Lorentz force. In Sec. \ref{sec:summary}, we compare the transport properties for 3D Weyl fermions, 2D massless Dirac fermions, and 3D conventional electrons. From Secs. \ref{sec:sigma-cal} through \ref{sec:sigma-c-cal}, we present detailed calculations for different contributions to the conductivity and magnetoconductivity.

\section{Model and method}\label{sec:model}

One of the low-energy descriptions of the
interacting Weyl semimetal is
\begin{eqnarray}\label{H}
H=\sum_{\mathbf{k},\nu}\psi_{\mathbf{k\nu}}^{\dag}[\nu\hbar v_{F}\boldsymbol{\sigma}\cdot(\mathbf{k}+\nu\mathbf{k}_{c})]\psi_{\mathbf{k\nu}}+\sum_{\mathbf{q}}\frac{V_{\mathbf{q}}}{2}\hat{\rho}_{\mathbf{q}}\hat{\rho}_{-\mathbf{q}},\nonumber\\
\end{eqnarray}
where $\psi_{\mathbf{k}\nu}^{\dag}=(\psi_{\mathbf{k}\nu\uparrow}^{\dag},\psi_{\mathbf{k}\nu\downarrow}^{\dag})$
is a two-component spinor operator with the valley index $\nu=\pm$
describing the opposite chirality and $\uparrow/\downarrow$ for the
spin index. The corresponding density operator is $\hat{\rho}_{\mathbf{q}}$=$\sum_{\nu,\mathbf{k}}\psi_{\mathbf{k}\nu}^{\dag}\psi_{\mathbf{k}+\mathbf{q},\nu}$.
$v_{F}$ is the Fermi velocity, $\hbar$ is the reduced Planck constant,
$\boldsymbol{\sigma}=(\sigma_{x},\sigma_{y},\sigma_{z})$ is the vector
of Pauli matrices, and $\pm\mathbf{k}_{c}$ are the two Weyl nodes. In
international unit $V_{\mathbf{q}}=e^{2}/\varepsilon q^{2}$ in 3D,
with $\varepsilon$ the dielectric constant. In realistic materials,
Weyl fermions are also perturbed by disorder $U(\mathbf{r})$. For mathematical convenience,
we assume the delta potential $U(\mathbf{r})=\sum_{i}u_{i}\delta(\mathbf{r}-\mathbf{R}_{i}),$where
$u_{i}$ measures the random potential at position $\mathbf{R}_{i}$,
and delta correlation between the impurities, $\langle U(\mathbf{r})U(\mathbf{r}')\rangle\sim\delta(\mathbf{r}-\mathbf{r}')$.

\begin{figure}[tbph]
\centering \includegraphics[width=0.45\textwidth]{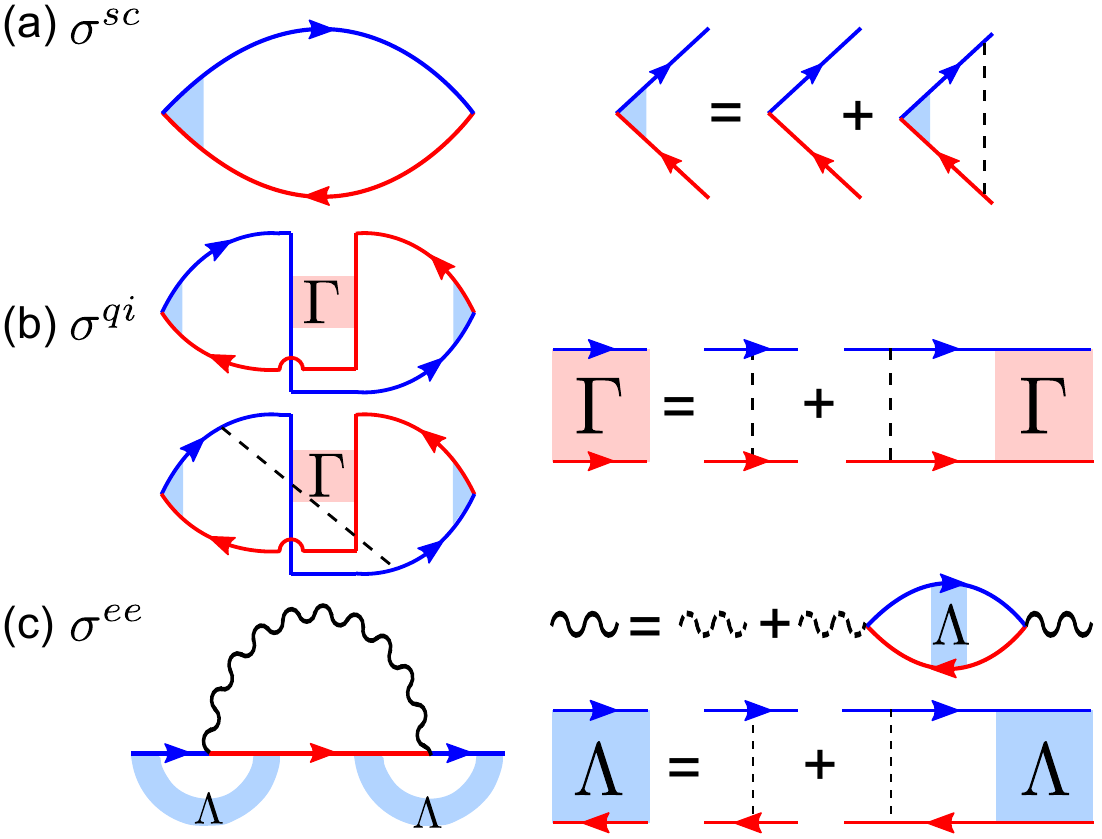}
\protect\caption{The Feynman diagrams \cite{McCann06prl,Altshuler80prl,Fukuyama80jpsj,Lee85rmp,Lu11prl,Shan12prb,Lu14prl}
for the conductivity of 3D Weyl semimetals, in the presence of disorder
(dashed lines) and electron-electron interaction (wavy lines). The
arrow lines are for Green's functions.}
\label{fig:diagram}
\end{figure}

We employ the Feynman diagram techniques to calculate the conductivity
in the presence of disorder and interaction (see Fig. \ref{fig:diagram}).
In this theoretical framework, the conductivity includes three dominant
parts, the semiclassical (Drude) conductivity \cite{Biswas14prb,Gorbar14prb} $\sigma^{sc}$ {[}Fig.
\ref{fig:diagram}(a){]}, the correction from the quantum interference
$\sigma^{qi}$ {[}Fig. \ref{fig:diagram}(b){]}, and the correction
from the interplay of electron-electron interaction and disorder scattering
$\sigma^{ee}$ {[}Fig. \ref{fig:diagram}(c){]}. We will first focus
on one valley, then move on to the multivalley case. Along an arbitrary
measurement direction, the Drude conductivity is found as $\sigma^{sc}=e^{2}N_{F}D$
(see Sec. \ref{sec:sc-cal} for the calculation), which satisfies
the Einstein relation. The density of states at the Fermi energy $E_{F}$
per valley $N_{F}=E_{F}^{2}/2\pi^{2}(v_{F}\hbar)^{3}$, the diffusion
coefficient $D=v_{F}^{2}\tau\eta_{v}/3$, with $\tau$ the total momentum relaxation time and the correction to velocity
by the ladder diagrams $\eta_{v}=3/2$ \cite{Garate12prb,Biswas14prb}.
We find that intervalley scattering can modify $\eta_{v}$ to $(3/2)/(1+\eta_{I})$,
where $\eta_{I}\in[0,1]$ measures the weight of the intervalley scattering
in the total scattering.

\section{Finite-temperature conductivity}\label{sec:sigma-temperature}

\subsection{Quantum interference and weak antilocalization}

\begin{figure}[tbph]
\centering \includegraphics[width=0.48\textwidth]{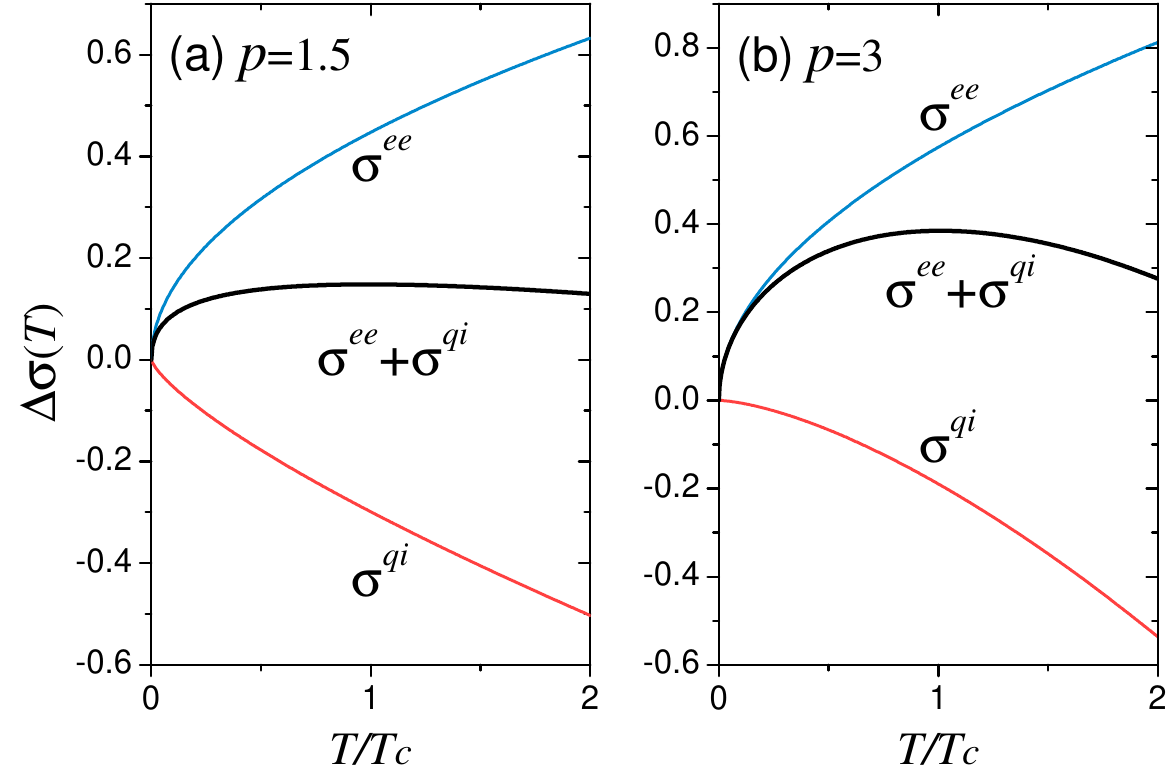}
\protect\caption{A schematic demonstration of the change of conductivity $\Delta\sigma$
as a function of temperature $T$. We choose $c_{ee}=c_{qi}$. $T_{c}$
is the critical temperature below which the conductivity drops with
decreasing temperature.}
\label{fig:sigma-T}
\end{figure}

According to the
classification of random ensembles \cite{Dyson62jmp}, systems with
time-reversal symmetry but broken spin-rotational symmetry are classified
into the symplectic class. A symplectic system is supposed to exhibit the weak antilocalization effect \cite{HLN80ptp}, when the quantum interference [see Fig. \ref{fig:diagram}(b)] corrects
the conductivity. Weyl fermions
in a single valley have the symplectic symmetry so the weak antilocalization effect is expected.
We find that the quantum interference correction
for one valley of Weyl fermions takes the form (detailed calculation
in Sec. \ref{sec:qi-cal})
\begin{eqnarray}
\sigma^{qi}(T)=\frac{e^{2}}{h}\frac{1}{\pi^{2}}(\frac{1}{\ell}-\frac{1}{\ell_{\phi}}),\label{sigma-qi}
\end{eqnarray}
where $e^{2}/h$ is the conductance quantum, $\ell$ is the mean free
path, and $\ell_{\phi}$ is the phase coherence length. This single-valley
result has exactly the same magnitude but opposite sign compared to
that for conventional 3D electrons (with dispersion $(\hbar k)^{2}/2m$)
per spin \cite{Lee85rmp}. With decreasing temperature, $\ell_{\phi}$
always increases as decoherence induced by inelastic scattering is
suppressed gradually. Therefore, $\sigma^{qi}$ will be enhanced when
lowering the temperature, literally giving a weak antilocalization contribution (see $\sigma^{qi}$ in Fig. \ref{fig:sigma-T}). The temperature dependence
of $\sigma^{qi}$ is from $\ell_{\phi}=CT^{-p/2}$ \cite{Thouless77prl},
where $C$ is a constant and $p$ depends only on dimensionality and
decoherence mechanisms thus does not distinguish conventional systems
and Weyl semimetals. In 3D, $p=3/2$ ($p=3$) if electron-electron
(electron-phonon) interaction is the decoherence mechanism in the disordered
limit \cite{Lee85rmp}. Also, because our calculation is in 3D, the
functional relationship is not logarithmic as that in quasi-2D \cite{Garate12prb,Lu11prb}.
For Weyl semimetals realized by breaking time-reversal symmetry \cite{Xu11prl}, the weak antilocalization may be suppressed by magnetism.

\subsection{Weak localization induced by the interplay of interaction and disorder}

Despite the signature
in magnetoconductivity, we will show that the weak antilocalization breaks
down in the presence of many-body interactions. The dominant interaction
correction to the conductivity is attributed to the one-loop Fock
(exchange interaction) self-energy dressed by Diffusons [see Fig. \ref{fig:diagram}(c)].
We find that, for Weyl fermions, this self-energy gives a correction
of the same form as that for conventional 3D electrons, upon a redefinition
of the parameters such as the diffusion coefficient $D$. Besides
the self-energy in Fig. \ref{fig:diagram}(c), there are three other
one-loop self-energies (see Fig. \ref{fig:one-loop}).
These four self-energies contribute to a correction to the conductivity
\begin{eqnarray}
\sigma^{ee}(T)\approx\frac{e^{2}}{h}(1-F)\sqrt{\frac{k_{B}T}{\hbar D}}\times0.195,
\end{eqnarray}
where $k_{B}$ is the Boltzmann constant and the screening factor $F$
is defined as the average of the interaction over the Fermi surface.
In 3D, $F=[\ln(1+x)]/x$ \cite{Lee85rmp}, where we find $x=8\pi^{2}v_{F}\hbar\varepsilon/e^{2}$
for Weyl fermions, with $\varepsilon$ the dielectric constant. By
definition, $F\in[0,1]$, as shown in Table \ref{tab:F} for several
popular candidates of Weyl semimetal. Therefore, $\sigma^{ee}$ decreases
with decreasing temperature following a law of $\sqrt{T}$, giving
a localization tendency.
Disorder is inevitable in realistic materials so here the interaction
is dressed by disorder, while in the clean limit the interaction
alone may give a linear-$T$ conductivity \cite{Goswami11prl,Hosur12prl}.
In the clean limit, a marginal Fermi liquid phase and generation of mass are found for Dirac semimetals \cite{Gonzalez14prb}. Also,
$F$ will be further corrected to $\widetilde{F}$ after including
the second-order interaction self-energies and interaction correction
to the disorder scattering \cite{Lee85rmp}. However $\widetilde{F}\sim F$
and $\widetilde{F}\le F$ (see Tab. \ref{tab:F} and Fig. \ref{fig:F}).
Later, we will see that there is always a localization tendency as
long as $1-F>0$, where the dominant 1 is contributed by the self-energy
in Fig. \ref{fig:diagram}(c). The interaction part $\sigma^{ee}$
also contributes to a negative magnetoconductivity, with a magnitude much smaller than $\delta\sigma^{qi}(B)$, this property is consistent with conventional
electrons \cite{Lee85rmp}.

\begin{table}[htbp]
\protect\caption{The dielectric constant $\varepsilon$ (in units of vacuum dielectric
constant), Fermi velocity $v_{F}$, and screening factor $F$ and
$\widetilde{F}$ (after the renormalization) for several candidates of
Weyl semimetals. Because of anisotropy, $v_{F}\hbar$ covers a wide
range in Bi$_{0.97}$Sb$_{0.03}$ and Cd$_{3}$As$_{2}$. }
\label{tab:F}\begin{ruledtabular} %
\begin{tabular}{ccccc}
 & $\varepsilon$  & $v_{F}\hbar$ {[}eV$\cdot$\AA {]}  & $F$  & $\widetilde{F}$ \tabularnewline
\hline
Bi$_{0.97}$Sb$_{0.03}$  & 100  & 1-10  & 0.09-0.01  & 0.09-0.01\tabularnewline
Refs. & [\onlinecite{Boyle60pr}]  & [\onlinecite{Golin68pr},\onlinecite{Liu95prb}]  &  & \tabularnewline
TlBiSSe  & 20  & 1.1  & 0.25  & 0.24 \tabularnewline
Refs. & [\onlinecite{Novak15prbr}]  & [\onlinecite{Novak15prbr}]  &  & \tabularnewline
Cd$_{3}$As$_{2}$  & 36-52  & 2-7  & 0.03-0.11  & 0.03-0.11 \tabularnewline
Refs. & [\onlinecite{Cisowski75jpss,JayGerin77ssc,Blom77}]  & [\onlinecite{Wang13prb},\onlinecite{Borisenko14prl}]  &  & \tabularnewline
\end{tabular}\end{ruledtabular}
\end{table}

\subsection{Competing weak antilocalization and localization}

Combining $\sigma^{qi}$ and $\sigma^{ee}$, the change of conductivity
with temperature for one valley of Weyl fermions can be summarized
as
\begin{eqnarray}
\Delta\sigma(T)=c_{ee}T^{1/2}-c_{qi}T^{p/2},
\end{eqnarray}
where $c_{ee}=0.195(1-F)\sqrt{k_{B}/\hbar D}$ and $c_{qi}=1/c\pi^{2}$
in units of $e^{2}/h$. This describes a competition between the interaction-induced
weak localization and interference-induced weak antilocalization, as shown in
Fig. \ref{fig:sigma-T} schematically. At higher temperatures, the
conductivity increases with decreasing temperature, showing a weak antilocalization
behavior. Below a critical temperature $T_{c}$, the conductivity
starts to drop with decreasing temperature, exhibiting a localization
tendency. From $\partial\Delta\sigma/\partial T=0$, the critical
temperature can be found as $T_{c}=\left(c_{ee}/p\cdot c_{qi}\right)^{2/(p-1)}$,
at which $\left.(\partial^{2}\Delta\sigma/\partial T^{2})\right|_{T_{c}}\approx(1-p)(p\cdot c_{qi}/4)\left(c_{ee}/p\cdot c_{qi}\right)^{(p-4)/(p-1)}$.
Because $c_{ee},c_{qi}>0$, this means as long as $p>1$, there is
always a critical temperature, below which the conductivity drops
with decreasing temperature. For known decoherence mechanisms in 3D,
$p$ is always greater than 1 \cite{Lee85rmp}. Now we estimate the
critical temperature $T_{c}$. Using $c_{ee}$ and $c_{qi}$, we arrive
at $T_{c}\approx\left[C(1-F)/(2p\sqrt{v_{F}\ell})\right]^{2/(p-1)}$,
which shows that $T_{c}$ increases with $C$ while decreases with
$F$, $p$, $v_{F}$, and $\ell$. With a set of typical parameters
$F=0.25\sim0.01$ and $v_{F}=10\sim1\times10^{5}$ m/s, as well as
$p=3\sim3/2$, $C=100\sim1000$ nm$\cdot$K$^{p/2}$, $\ell=100\sim10$
nm in disordered metals, we find that $T_{c}\approx0.4\sim10^{6}$
K. Please note that our calculation is not justified at high temperatures, but in this way we show that the
localization tendency is experimentally accessible in disordered
Weyl semimetals. For Cd$_{3}$As$_{2}$ with extremely high mobility,
the recent experiment \cite{Liang15nmat} demonstrated that the mean
free path $\ell$ is well above 1 $\mu$m, yielding a $T_{c}$ well
below those achievable (10 mK) in most laboratories. To summarize
the transport properties of a single valley of Weyl fermions, Table
\ref{tab:2D-3D} compares them with those of 2D massless Dirac fermions
and 3D conventional electrons.

\section{Magnetoconductivity}\label{sec:MC}

\begin{figure}[tbph]
\centering
\includegraphics[width=0.49\textwidth]{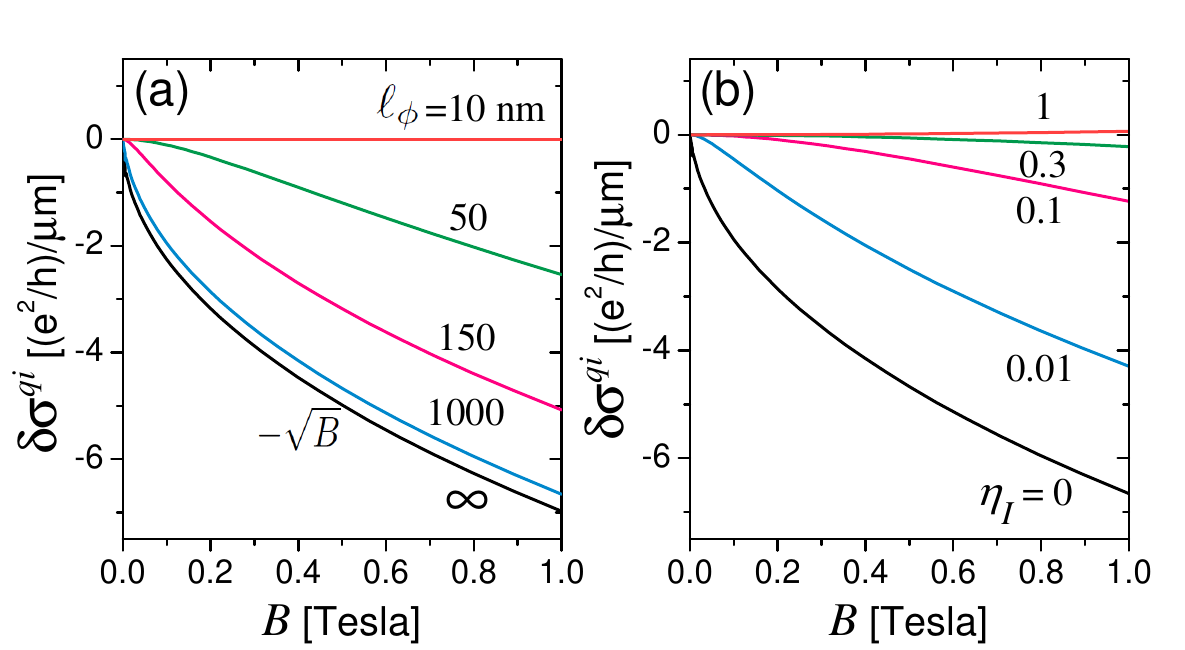}
\includegraphics[width=0.48\textwidth]{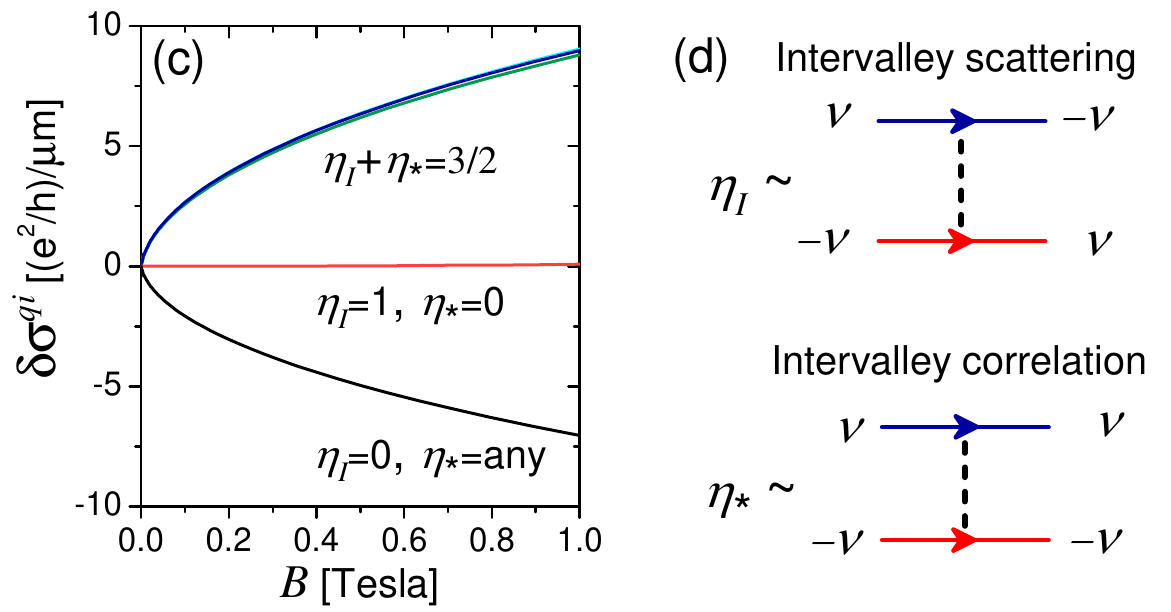}
\protect\caption{The magnetoconductivity $\delta\sigma^{qi}(B)$ for different phase
coherence length $\ell_{\phi}$ at $\eta_{I}=\eta_{*}=0$ (a), for
different $\eta_{I}$ at $\eta_{*}=0$ (b), and for different $\eta_{I}$
at finite $\eta_{*}$ (c). Parameters: $\ell=10$ nm and $\ell_{\phi}=1000$
nm in (b) and (c). (d) The diagrams show the difference between $\eta_{I}$
and $\eta_{*}$, with $\eta_{I}$ related to the intervalley scattering
and $\eta_{*}$ measuring the intervalley correlation of intravalley
scattering. The dashed lines represent the correlation of two scattering
processes. $\nu=\pm$ is the valley index.}
\label{fig:MC}
\end{figure}

\subsection{The $-\sqrt{B}$ magnetoconductivity of a single valley of Weyl fermions}\label{sec:mc-wal}

Because of its quantum interference origin, $\sigma^{qi}$ in Eq.
(\ref{sigma-qi}) can be suppressed by a magnetic field, giving rise
to a magnetoconductivity $\delta\sigma^{qi}(B)\equiv\sigma^{qi}(B)-\sigma^{qi}(0)$,
where
\begin{eqnarray}
 &  & \sigma^{qi}(B)=\frac{2e^{2}}{h}\int_{0}^{1/\ell}\frac{dx}{(2\pi)^{2}}\nonumber \\
 &  & \times\left[\psi\left(\frac{\ell_{B}^{2}}{\ell^{2}}+\ell_{B}^{2}x^{2}+\frac{1}{2}\right)-\psi\left(\frac{\ell_{B}^{2}}{\ell_{\phi}^{2}}+\ell_{B}^{2}x^{2}+\frac{1}{2}\right)\right]
\end{eqnarray}
for one valley of Weyl fermions,
with $\psi$ the digamma function and $\ell_{B}\equiv\sqrt{\hbar/4eB}$
the magnetic length. The magnetic field $B$ is applied along arbitrary
directions (we have checked that $\sigma_{xx}^{qi}=\sigma_{zz}^{qi}$).
As $B\rightarrow\infty$, $\delta\sigma^{qi}$ saturates following
a $1/B$ dependence. As $B\rightarrow0$, $\delta\sigma^{qi}$ is
proportional to $-\sqrt{B}$ for $\ell_{\phi}\gg\ell_{B}$ or at low
temperatures, and $\delta\sigma^{qi}\propto-B^{2}$ for $\ell_{\phi}\ll\ell_{B}$ at high temperatures.
$\ell_{B}$ can be evaluated approximately as 12.8 nm$/\sqrt{B}$
with $B$ in Tesla. Usually below the liquid helium temperature, $\ell_{\phi}$
can be as long as hundreds of nanometers to one micrometer, much longer than
$\ell_{B}$ which is tens of nanometers between 0.1 and 1 Tesla. Therefore,
the $-\sqrt{B}$ magnetoconductivity at low temperatures and small
fields serves as a signature for the weak antilocalization of 3D Weyl fermions.
Fig. \ref{fig:MC}(a) shows $\delta\sigma^{qi}(B)$ of two valleys
of Weyl fermions in the absence of intervalley scattering. For long $\ell_{\phi}$,
$\delta\sigma^{qi}(B)$ is negative and proportional to $\sqrt{B}$,
showing the signature of the weak antilocalization of 3D Weyl fermions. This
$-\sqrt{B}$ dependence agrees well with the experiment \cite{Kim13prl,Kim14prb},
and we emphasize that it is obtained from a complete diagram calculation
with only two parameters $\ell$ and $\ell_{\phi}$ of physical meanings.
As $\ell_{\phi}$ becomes shorter, a change from $-\sqrt{B}$ to $B^{2}$
is evident. $\delta^{qi}(B)$ vanishes at $\ell_{\phi}=\ell$ as the
system quits the quantum interference regime. Also, it is known that
the chiral anomaly could give a positive magnetoconductivity \cite{Son13prb,Kim14prb,Burkov14prl-chiral,Gorbar14prb},
competing with the negative magnetoconductivity from the weak antilocalization.
This chiral-anomaly part, because of its $B^{2}$ dependence, will
always be overwhelmed by the $-\sqrt{B}$ weak antilocalization part at weak magnetic fields. At high fields, the chiral anomaly may become dominant.

\subsection{Weak localization induced by inter-valley effects}

Now we come to consider the effects
of intervalley scattering and correlation. We will focus on the quantum
interference part and magnetoconductivity [see Sec. \ref{sec:qi-c-mc} for the expressions of $\sigma^{qi}$ and $\delta\sigma^{qi}(B)$
in the presence of intervalley scattering and correlation], because
we find that the interaction brings a negligible valley-dependent
effect. Two dimensionless parameters are defined for the inter- and
intravalley scattering: $\eta_{*}\propto\langle U_{\mathbf{k},\mathbf{k}'}^{++}U_{\mathbf{k}',\mathbf{k}}^{--}\rangle$
measuring the correlation between intravalley scattering and $\eta_{I}\propto\langle U_{\mathbf{k},\mathbf{k}'}^{+-}U_{\mathbf{k}',\mathbf{k}}^{-+}\rangle$
measuring the weight of intervalley scattering , where $U_{\mathbf{k},\mathbf{k}'}^{\nu,\nu'}$
is the scattering matrix element. Figure \ref{fig:MC}(d)
schematically shows the difference between $\eta_{*}$ and $\eta_{I}$.
As shown in Fig. \ref{fig:MC}(b),
with increasing $\eta_{I}$, the negative $\delta\sigma^{qi}$ is
suppressed, where $\eta_{I}\rightarrow1$ means strong intervalley
scattering while $\eta_{I}\rightarrow0$ means vanishing intervalley
scattering. Furthermore, Fig. \ref{fig:MC}(c) shows that the magnetoconductivity
can turn to positive when $\eta_{I}+\eta_{*}=3/2$.
Remember that the negative $\delta\sigma^{qi}(B)$ in Fig. \ref{fig:MC}(b)
is related to the increasing $\sigma^{qi}(T)$ with decreasing $T$
in Fig. \ref{fig:sigma-T}, as two signatures of the weak antilocalization.
Similarly, the positive $\delta\sigma^{qi}(B)$ in Fig. \ref{fig:MC}(c)
corresponds to a suppressed $\sigma^{qi}$ with decreasing temperature,
i.e., a localization tendency. This localization is attributed to the strong intervalley coupling which recovers spin-rotational symmetry (now the spin space is complete for a given momentum), then the system goes to the orthogonal class \cite{Dyson62jmp,HLN80ptp,Suzuura02prl}.
Therefore, we show that the combination
of strong intervalley scattering and correlation will strengthen the
localization tendency in disordered Weyl semimetals.

\subsection{Comparison with magnetoconductivity in experiments}

\begin{table}[htbp]
\caption{Comparison between the classical magnetoconductivity (MC) induced by the Lorentz force $\delta \sigma^C(B)$, semiclassical MC induced by the chiral anomaly $\delta\sigma^A(B)$ \cite{Son13prb,Burkov14prl-chiral}, and weak antilocalization MC induced by the quantum interference $\delta \sigma^{qi}(B)$, in their dependences on magnetic field $B$, temperature $T$, Fermi wave vector $k_F$, and mean free path $\ell$, in the limit $\eta_I\rightarrow 0$.  }
\label{tab:MC}%
\begin{ruledtabular}
\begin{tabular}{cccc}
Dependence & $\delta \sigma^C$   & $\delta\sigma^A$ &$\delta \sigma^{qi} (\eta_I\rightarrow 0)$   \\ \hline
$B$ & $-B^2$ & $B^2$  &  $-\sqrt{B}$    \\
$B$ direction & $\perp$ & $||$  &  Any    \\
$T$ & No  & No & Suppressed with increasing $T$   \\
$k_F$ & No  & $1/k_F^2$ & No   \\
$\ell$ & $\ell^3$  & $\ell $ & Decreases with increasing $\ell$   \\
$\eta_I$ & No & $ 1/\eta_I$ &  Suppressed with increasing $\eta_I$
 \end{tabular}
\end{ruledtabular}
\end{table}

To compare with experiments, besides the magnetoconductivity $\delta\sigma^{qi}(B)$ arising from the weak antilocalization in Sec. \ref{sec:mc-wal}, two more contributions to the total magnetoconductivity have to be taken into account. One is the
classical negative magnetoconductivity due the cyclotron motion of electron driven by the Lorentz force in perpendicular magnetic fields and is given by $\delta \sigma^C= -\sigma^{sc} \mu^2 B^2 $ \cite{Datta1997}, where for the Weyl fermion the mobility is given by $\mu=ev_F \tau \eta_v /\hbar k_F$, then (see Sec. \ref{sec:sigma-c-cal} for details)
\begin{eqnarray}\label{MC-c}
\delta\sigma^{C}(B) =-\frac{e^2}{h}\frac{\sqrt{3}\eta_v^{3/2}}{16\pi}\frac{\ell^3}{\ell_B^4}.
\end{eqnarray}
This part arises only in a perpendicular field and is not a function of $k_F$. It becomes dominant for long $\ell$, i.e., in high-mobility and clean samples.

The other semiclassical magnetoconductivity is from the chiral anomaly, which arises because of the nontrivial Berry curvature carried by Weyl fermions, and it can give a magnetic field dependent correction to the velocity and Drude conductivity. An explicit form of $\delta\sigma^{A}(B)$ has been derived by Son and Spivak \cite{Son13prb} and Burkov \cite{Burkov14prl-chiral}. For example, according to Burkov \cite{Burkov14prl-chiral}
\begin{eqnarray}
\delta\sigma^{A}(B) &=&  \frac{e^4B^2\tau_a}{4\pi^4g(E_F)}
\end{eqnarray}
where $ \hbar =1$, $g(E_F)=2N_F$. $\tau_a$ is referred to as the axial relaxation time, which is supposed to be an independent parameter. Here, we use the intervalley scattering time for the axial relaxation time. In terms of the notations used in this work
\begin{eqnarray}\label{MC-ca}
\delta\sigma^{A}(B)=\frac{e^2}{h } \frac{ \ell }{\ell_B^4 } \frac{1}{ k_F^2} \frac{\sqrt{2(1+\eta_I)}}{32\pi \eta_I }.
\end{eqnarray}
Here $\ell_{B}\equiv \sqrt{\hbar/4eB }$. The Berry curvature correction may also be the reason for some anomalous magnetoconductivity in topological insulators \cite{Wang12nr}.

Including the three contributions, now the total magnetoconductivity is
\begin{eqnarray}\label{MC-para}
\delta\sigma_{||}(B) &=& \delta\sigma^{qi}(B) + \delta \sigma^{A}(B)
\end{eqnarray}
when the current is parallel to the magnetic field, and
\begin{eqnarray}\label{MC-perp}
\delta\sigma_\perp(B) &=& \delta\sigma^{qi}(B) +\delta \sigma^C(B)
\end{eqnarray}
when the current is perpendicular to the magnetic field.

In Table \ref{tab:MC}, we compare these three different magnetoconductivity. Please note that, $\delta\sigma^C$ dominates in clean samples because it is proportional to $\ell^3$ while $\delta\sigma^A$ dominates near Weyl nodes because it is proportional to $1/k_F^2$, and $\delta\sigma^{qi}$ appear only at low temperatures.

\begin{figure}[tbph]
\centering
\includegraphics[width=0.48\textwidth]{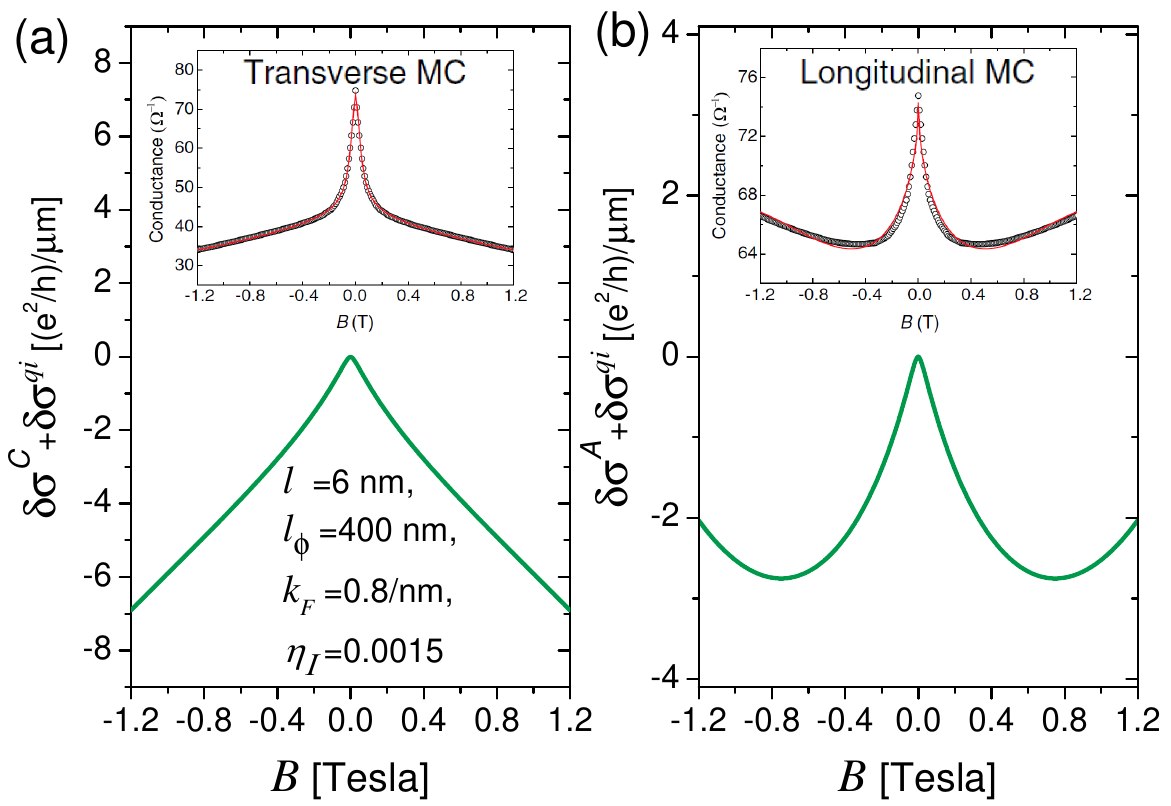}
\caption{Comparison with the experimental magnetoconductivity by Kim \emph{et al}. (insets) \cite{Kim13prl}. The transverse magnetoconductivity (MC) $\delta\sigma_\perp(B)=\delta\sigma^C+\delta\sigma^{qi}$ and longitudinal MC $\delta\sigma_{||}(B)=\delta\sigma^A+\delta\sigma^{qi}$ are defined when the current is perpendicular and parallel to the magnetic field, respectively.
The classical MC induced by the Lorentz force $\delta \sigma^C(B)$ is given in Eq. (\ref{MC-c}). The semiclassical MC induced by the chiral anomaly $\delta\sigma^A(B)$ \cite{Son13prb,Burkov14prl-chiral} is in Eq. (\ref{MC-ca}). The delocalization MC induced by the quantum interference $\delta \sigma^{qi}(B)$ is in Sec. (\ref{sec:qi-c-mc}). }
\label{fig:Kim}
\end{figure}

In Fig. \ref{fig:Kim}, we use Eqs. (\ref{MC-para}) and (\ref{MC-perp}) to reproduce the magnetoconductivity measured by Kim \emph{et al}. in Bi$_{0.03}$Sb$_{0.97}$ \cite{Kim13prl}. The main features (e.g., the transverse MC is several times of the longitudinal MC, those inflection points in MC) in both the transverse and longitudinal magnetoconductivity can be recovered simultaneously within a set of parameters comparable to those in relevant materials. Our parameters (e.g., mean free path, Fermi wave vector) are of physical meanings. Figure \ref{fig:Kim} (a) is always negative because both $\delta\sigma^C$ and $\delta\sigma^{qi} $ are negative. The competition between $\delta\sigma^{qi}\propto -\sqrt{B}$ and $\delta\sigma^A\propto B^2$ leads to the inflection in Fig. \ref{fig:Kim} (b).

\begin{widetext}

\section{Comparison of Weyl fermions, 2D Dirac fermions, and 3D conventional electrons}\label{sec:summary}

To summarize, we compare the transport properties for single valley of Weyl Fermions, 2D massless Dirac fermions, and 3D conventional electrons, in Table \ref{tab:2D-3D}.

\begin{table}[htbp]
\caption{Comparison between a single valley of 2D massless Dirac fermions \cite{Suzuura02prl,McCann06prl}, a single
valley of 3D Weyl fermions, and a single band of 3D conventional electrons \cite{Altshuler80prl,Fukuyama80jpsj}.
$k_{F}$ is the Fermi wave vector, $m$ is the effective mass, $v_{F}$ is the constant Fermi
velocity of Dirac and Weyl fermions, $\tau$ is the total momentum relaxation time, and $\eta_{v}$ is the correction to the velocity (Sec. \ref{sec:sc-cal}). $\eta_{I}\in[0,1]$
measures the weight of intervalley scattering. $F$ is the screening
factor of interaction (Sec. \ref{sec:F-cal}). $\varepsilon$ is the dielectric
constant. $\sigma^{ee}$ is the conductivity correction from the interplay
of electron-electron interaction and disorder scattering. $\sigma^{qi}$
is the conductivity correction from the quantum interference. $\delta\sigma^{qi}(B)$
is the small-field magnetoconductivity from $\sigma^{qi}$ when $\ell_{\phi}\gg\ell_{B}$. $p$ is the exponent in the temperature dependence of the phase coherence length \cite{Thouless77prl} $\ell_\phi\sim T^{-p/2}$ due to electron-electron interaction (EEI) and electron-phonon (E-Ph) interaction in disordered metals \cite{Lee85rmp}.}
\label{tab:2D-3D}%
\begin{ruledtabular}
\begin{tabular}{cccc}
 & 2D massless Dirac & 3D Weyl  & 3D Conventional  \\
\hline \\
Dispersion $E(\mathbf{k})$ & $ \pm \gamma\sqrt{k_x^2+k_y^2}$ &  $ \pm \gamma \sqrt {k_x^2+k_y^2+k_z^2}$   & $(\hbar^2/2m) (k_x^2+k_y^2+k_z^2)$
\\
Density of states $N(E)$ & $E/ 2\pi (v_F\hbar)^2 $ &  $E^2/ 2\pi^2 (v_F\hbar)^3 $  & $(1/2\pi)^2 (2m/\hbar^2)^{3/2} \sqrt{E}$  \\
Carrier density per valley & $k_F^2/ 4\pi $  &  $k_F^3/ 6\pi^2 $  &  $k_F^3/ 6\pi^2 $  \\
Mobility $\mu$ &  $ev_F\tau \eta_v / \hbar k_F $ &  $ev_F\tau \eta_v / \hbar k_F $  & $e\tau /m   $  \\
Diffusion coefficient $D$ & $ v_F^2\tau\eta_v/2$ &  $ v_F^2\tau\eta_v/3$  & $ (\hbar k_F/m)^2\tau /3$ \\
Velocity correction $\eta_v$ & 2 \cite{Shon98jpsj} &  3/2 \cite{Garate12prb,Biswas14prb}  & 1 \\
  &   &  $(3/2)/(1+\eta_I)$   &   \\
$\eta_H$ & -1/4 \cite{McCann06prl} &  -1/6 \cite{Garate12prb}  & 0 \\
Screening factor $F$ &  $(2/\pi)(\arctan\sqrt{1/x^2-1})/\sqrt{1-x^2}$ & $[\ln(1+x)]/x$ & $[\ln(1+x)]/x$ \\
& $x= 8\pi \varepsilon v_F\hbar/e^2$  & $x= 8\pi^2 \varepsilon v_F\hbar /e^2$ & $x=  8\pi^2 \varepsilon \hbar^2 k_F/m e^2  $ \\
$\delta\sigma^{qi}(B\rightarrow0)\propto$  & $-B$  & $-\sqrt{B}$  & $\sqrt{B}$ \cite{Kawabata80jpsj}\tabularnewline
$\sigma^{ee}(T)\propto$  & $\ln T$ \cite{Lu14prl,LiuHC14an} & $\sqrt{T}$  & $\sqrt{T}$ \cite{Altshuler80prl,Fukuyama80jpsj} \tabularnewline
$\sigma^{qi}(T)\propto$  & $-\ln T$ \cite{Suzuura02prl,McCann06prl}  & $-T^{p/2}$  & $T^{p/2}$ \cite{Altshuler80prl,Fukuyama80jpsj}\tabularnewline
$p$ (EEI) \cite{Lee85rmp} & 1   & 3/2 & 3/2  \\
$p$ (E-Ph) \cite{Lee85rmp} & 3   & 3 & 3 \\
\end{tabular}
\end{ruledtabular}
\end{table}

\end{widetext}

\section{The Calculation of the conductivity}\label{sec:sigma-cal}

Throughout the work, we will only focus on the conductivity of the conduction bands. The valence bands have the same properties. The eigen energies of the conduction bands in the two valleys are degenerate
\begin{eqnarray}
E_{\mathbf{k} }= v_F\hbar  k=v_F\hbar \sqrt{k_x^2+k_y^2+k_z^2},
\end{eqnarray}
where $\mathbf{k}$ is measured from each Weyl node. The spinor wave function of the conduction band in valley $+$ is
\begin{eqnarray}\label{c+}
|+,\mathbf{k}\rangle =\left(
  \begin{array}{c}
    \cos\frac{\theta}{2} \\
    \sin\frac{\theta}{2} e^{i\varphi} \\
  \end{array}
\right)\frac{e^{i\mathbf{k}\cdot\mathbf{r}}}{\sqrt{\Omega}},
\end{eqnarray}
where $\theta$ and $\varphi$ are the wave vector angles, $\tan\varphi\equiv k_y/k_x $, $\cos\theta\equiv k_z/k$, and $\Omega$ is the volume. In valley $-$,
the wave function of the conduction band can be found as ($\theta\rightarrow \pi-\theta$ and $\varphi\rightarrow \pi+\varphi $),
\begin{eqnarray}\label{c-}
|-,\mathbf{k}\rangle =\left(
  \begin{array}{c}
    \sin\frac{\theta}{2} \\
    -\cos\frac{\theta}{2} e^{i\varphi} \\
  \end{array}
\right)\frac{e^{i\mathbf{k}\cdot\mathbf{r}}}{\sqrt{\Omega}}.
\end{eqnarray}

The total conductivity has three dominant parts
\begin{eqnarray}
\sigma &=& \sigma^{sc}+\sigma^{qi}+\sigma^{ee}.
\end{eqnarray}
$\sigma^{sc}$ is the semiclassical conductivity (Sec. \ref{sec:sc-cal}), $\sigma^{qi}$ is the correction from the quantum interference (Sec. \ref{sec:qi-cal}), and $\sigma^{ee}$ is the correction from the interplay of electron-electron interaction and disorder scattering (Sec. \ref{sec:ee-cal}).

\section{Semiclassical (Drude) conductivity}\label{sec:sc-cal}

\begin{figure}[tbph]
\centering
\includegraphics[width=0.48\textwidth]{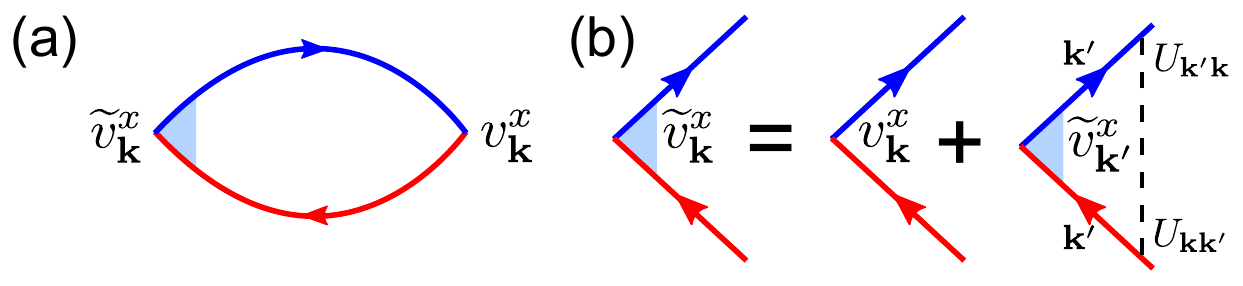}
\caption{(a) The Feynman diagram for the semiclassical (Drude) conductivity $\sigma^{sc}$.
(b) The diagram for the vertex correction to the velocity \cite{Shon98jpsj}. $v$ is the velocity. The arrow lines are for Green's functions. The dashed lines are for disorder scattering ($U$). Replace $x$ by $ z$ for the conductivity along the $z$ direction. }
\label{fig:Drude}
\end{figure}

The semiclassical (Drude) conductivity can be found as [see Fig. \ref{fig:Drude}(a)]
\begin{eqnarray}\label{sigma-sc-def}
\sigma^{sc}
= \frac{e^2\hbar}{2\pi}
\sum_{\mathbf{k}}  v^i_{\mathbf{k}} G^{\mathrm{R}}_{\mathbf{k}}
G^{\mathrm{A}}_{\mathbf{k}}\widetilde{v}^i_{\mathbf{k}},
\end{eqnarray}
where $i=x $ or $z$, $G^{R/A}$ is the retarded/advanced Green's function, $v^i_{\mathbf{k}}=(1/\hbar)\partial_i /\partial E_{\mathbf{k}}$ is the velocity, and $\widetilde{v}^i_{\mathbf{k}}$ is the corrected velocity by the disorder scattering [see Fig. \ref{fig:Drude} (b)].
The retarded (R) and advanced (A) Green's functions
\begin{eqnarray}
G^{\mathrm{R/A}}_{\mathbf{k}}(\omega)
&=&\frac{1}{\omega -\xi_{\mathbf{k}}\pm i\frac{\hbar}{2\tau}},
\end{eqnarray}
where $\xi_\mathbf{k}=E_{\mathbf{k}}-E_F$. The total scattering time (or total momentum relaxation time) $\tau$ is defined as
\begin{eqnarray}
\frac{1}{\tau}\equiv \frac{1}{\tau_0}+\frac{1}{\tau_I},
\end{eqnarray}
where the intra- and inter-valley scattering times are given by
\begin{eqnarray}
\frac{1}{\tau_0}
&\equiv & \frac{2\pi}{\hbar} \sum_{\mathbf{k}'}\langle |U^{++}_{\mathbf{k},\mathbf{k}'}|^2\rangle\delta(E_F-\xi_{\mathbf{k}'})  =\frac{2\pi}{\hbar }N_F\frac{n u_0^2}{2},\nonumber\\
\frac{1}{\tau_I}&\equiv &\frac{2\pi}{\hbar}\sum_{\mathbf{k}'}
\langle|U^{+-}_{\mathbf{k},\mathbf{k}'}|^2\rangle
\delta(E_F-\xi_{\mathbf{k}'})=\frac{2\pi}{\hbar} N_F \frac{nu_I^2}{2},
\end{eqnarray}
$n$ is the impurity density, $u_0$ and $u_I$ measure the strength for the intra- and inter-valley scattering, respectively. $N_F=E_F^2/2\pi^2(v_F\hbar)^3$ is the density of states per valley.
$U_{\mathbf{k},\mathbf{k}'}^{++}\equiv \langle +,\mathbf{k}|U(\mathbf{r})|+,\mathbf{k}'\rangle $ and $U_{\mathbf{k},\mathbf{k}'}^{+-}\equiv \langle +,\mathbf{k}|U(\mathbf{r})|-,\mathbf{k}'\rangle $ are the intravalley and intervalley scattering matrix elements, respectively, and
\begin{eqnarray}\label{U++U+-}
U^{++}_{\mathbf{k}, \mathbf{k}'}
&=&  \sum_i  u_i     e^{i(\mathbf{k}'-\mathbf{k}) \cdot \mathbf{R}_i}
[aa'+bb'e^{i(\varphi'-\varphi)}],\nonumber\\
U^{+-}_{\mathbf{k}, \mathbf{k}'}
&=&  \sum_i  u_i e^{i(\mathbf{k}'-\mathbf{k}) \cdot \mathbf{R}_i}
[a b'-b a'e^{i(\varphi'-\varphi)}],
\end{eqnarray}
with $a\equiv \cos(\theta/2)$ and $b\equiv \sin(\theta/2)$.

The correction to the velocity can be found from the iteration equation [see Fig. \ref{fig:Drude}(b)]
\begin{eqnarray}\label{vxiteration}
\widetilde{v}^i_{\mathbf{k}}&=&v^i_{\mathbf{k}}
+\sum_{\mathbf{k}'} G^{\mathrm{R}}_{\mathbf{k}'}
G^{\mathrm{A}}_{\mathbf{k}'}
\langle
U_{\mathbf{k},\mathbf{k}'}
U_{\mathbf{k}',\mathbf{k}}\rangle
\widetilde{v}^i_{\mathbf{k}'}.
\end{eqnarray}
In polar coordinates,
$v^x_\mathbf{k}=v_F\sin\theta\cos\varphi$ and $v_{\mathbf{k}}^z=v_F\cos\theta$,
\begin{eqnarray}
\int_0^\infty (k')^2\frac{dk'}{2\pi}
G^{\mathrm{R}}_{\mathbf{k}'}
G^{\mathrm{A}}_{\mathbf{k}'}
&\approx & \frac{2\pi^2 N_F \tau}{\hbar},
\end{eqnarray}
and
\begin{eqnarray}
&& \langle U_{\mathbf{k},\mathbf{k}'}U_{\mathbf{k}' ,\mathbf{k} }\rangle =\langle U^{++}_{\mathbf{k} ,\mathbf{k}'}U^{++}_{\mathbf{k}',\mathbf{k}}\rangle+\langle U^{+-}_{\mathbf{k},\mathbf{k}' }U^{-+}_{\mathbf{k}',\mathbf{k}}\rangle\nonumber\\
&&\approx  \frac{\hbar}{2\pi N_F\tau}[1+(1-2\eta_I)\nonumber\\
&&\times(\cos\theta \cos\theta'+\sin\theta \sin\theta'\cos(\varphi-\varphi'))],
\end{eqnarray}
where $\eta_I\equiv \tau/\tau_I$ and $1/\tau\equiv 1/\tau_0+1/\tau_I$, and $\tau_0$ and $\tau_I$ are the intravalley and intervalley scattering times, respectively. So $\eta_I\in[0,1]$ measures the relative strength of intervalley scattering. By assuming $\widetilde{v}^x_{\mathbf{k}}=\eta_v v_\mathbf{k}^{x}$ and $\widetilde{v}^z_{\mathbf{k}}=\eta_v v_\mathbf{k}^{z}$, and put them into the iteration equation for the velocity,
one can readily find that for either the velocity along $x$ or $z$ direction
\begin{eqnarray}
\eta_v &=& \frac{3}{2(1+\eta_I)}.
\end{eqnarray}

Finally, we found that for either $x$ and $z$ direction,
\begin{eqnarray}\label{drude}
\sigma^{sc}
&=& e^2  N_F \frac{1}{3}v_F^2\tau\eta_v,
\end{eqnarray}
where the density of states per valley $N_F=E_F^2/2\pi^2(v_F\hbar)^3$.
It satisfies the Einstein relation
\begin{eqnarray}\label{Einstein}
\sigma^{sc}
&=&  e^2 N_F D,
\end{eqnarray}
with the diffusion coefficient $D\equiv  v_F^2 \tau \eta_v/d$, where $d=3$ for three dimensions. Usually, $\tau\eta_v$ is referred to as the transport time.
Later, we will show that $D$ can also be derived from the calculation of the Diffuson (Sec. \ref{sec:Diffuson-cal}).

In terms of the mean free path $\ell\equiv \sqrt{D\tau}$ and Fermi wave vector $k_F$,
\begin{eqnarray}\label{Drude-L-kF}
\sigma^{sc}
&=&  \frac{e^2}{h}  \frac{k_F^2\ell }{\pi}\sqrt{\frac{\eta_v}{3}}= \frac{e^2}{h}  \frac{k_F^2\ell }{\pi\sqrt{ 2(1+\eta_I) }}.
\end{eqnarray}

\section{Conductivity correction from quantum interference}\label{sec:qi-cal}

The total conductivity from the quantum interference has two parts
\begin{eqnarray}
\sigma^{qi} &=& 2\times \sigma^{qi}_0+\sigma^{qi}_I.
\end{eqnarray}
$\sigma^{qi}_0$ is from the intravalley Cooperons (Sec. \ref{sec:qi-intra-cal}) and
$\sigma^{qi}_I$ is from the intervalley Cooperons (Sec. \ref{sec:qi-inter-cal}).

\subsection{Conductivity correction from intravalley Cooperons}\label{sec:qi-intra-cal}

\begin{figure}[tbph]
\centering
\includegraphics[width=0.48\textwidth]{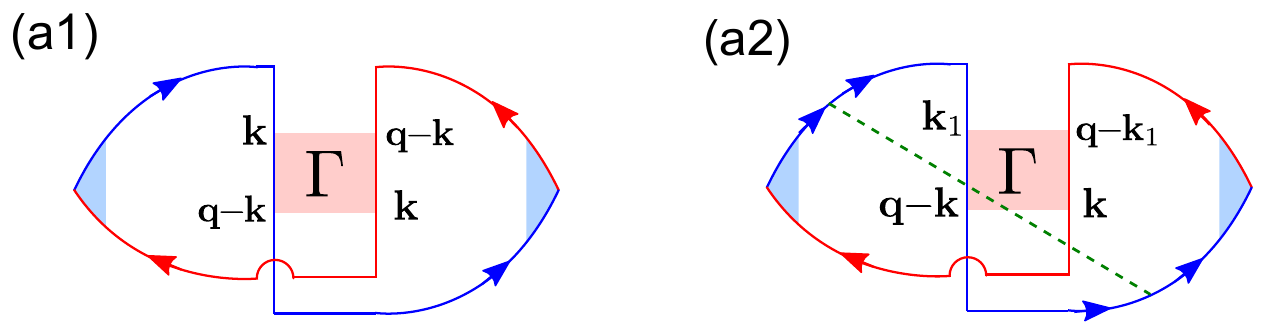}
\caption{The Feynman diagrams for the quantum interference correction to the conductivity that take into account the Cooperons from only the intravalley scattering. These diagrams give $\sigma^{qi}_0$. }
\label{fig:Hikami-boxes-intra}
\end{figure}
The conductivity contribution from the intravalley Cooperons is given by (see Fig. \ref{fig:Hikami-boxes-intra})
\begin{eqnarray}\label{sigma-qi-0-def}
\sigma^{qi}_0= \sigma_{a1}+2\times \sigma_{a2} ,
\end{eqnarray}
where
\begin{eqnarray}
\sigma_{a1} &=& \frac{e^2\hbar}{2\pi}\sum_{\mathbf{q}}\Gamma_{\mathbf{k},\mathbf{q}-\mathbf{k}}\sum_{\mathbf{k}}
G^R_{\mathbf{k}} \widetilde{v}^x_\mathbf{k}
G^A_{\mathbf{k}}
G^R_{\mathbf{q-k}}
\widetilde{v}^x_\mathbf{q-k}
G^A_{\mathbf{q-k}}, \nonumber\\
\sigma_{a2}
&=& \frac{e^2  \hbar}{2\pi} \sum_{\mathbf{q}}
\Gamma_{\mathbf{k}_1,\mathbf{q}-\mathbf{k}}
 \sum_{\mathbf{k}}\sum_{\mathbf{k}_1}
\widetilde{v}^x_{\mathbf{k}}\widetilde{v}^x_{\mathbf{q}-\mathbf{k}_1}
G^R_{\mathbf{k}}G^R_{\mathbf{k}_1}G^R_{\mathbf{q}-\mathbf{k}}
\nonumber\\
&&\times G^R_{\mathbf{q}-\mathbf{k}_1}G^A_{\mathbf{k}} G^A_{ \mathbf{q}-\mathbf{k}_1}\langle U_{\mathbf{k},\mathbf{k}_1}
U_{\mathbf{q}-\mathbf{k},\mathbf{q}-\mathbf{k}_1}\rangle.
\end{eqnarray}
We also find that the ratio of the dressed to bare Hikami boxes is
\begin{eqnarray}
\eta_H\equiv \frac{\sigma_{a2}}{\sigma_{a1}}=-\frac{1}{6},
\end{eqnarray}
consistent with that by Garate and Glazman \cite{Garate12prb}.

\begin{figure}[tbph]
\centering
\includegraphics[width=0.3\textwidth]{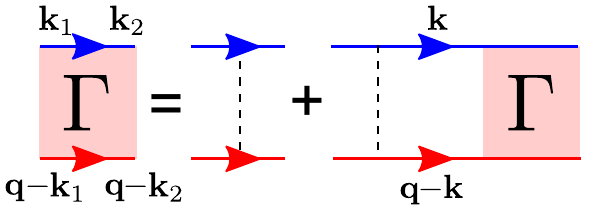}
\caption{The Feynman diagram of the Bethe-Salpeter equation for the intravalley Cooperons. The valley index is suppressed because the valley is conserved during the scattering.}
\label{fig:Cooperon-intra}
\end{figure}

In polar coordinates, the intra-valley Cooperon can be found by the Bethe-Salpeter equation (see Fig. \ref{fig:Cooperon-intra})
\begin{eqnarray}
\Gamma_{\mathbf{k}_1,\mathbf{k}_2}
&=&\Gamma_{\mathbf{k}_1,\mathbf{k}_2}^0
+\int_0^{2\pi}\frac{d\varphi}{2\pi}
\int_0^\pi \frac{d\theta\sin\theta}{2\pi}\int_0^\infty \frac{dk k^2}{2\pi}\nonumber\\
&&\times \Gamma_{\mathbf{k}_1,\mathbf{k}}^0 \mathcal{G}_\mathbf{k}^{i\epsilon_n}
\mathcal{G}_{\mathbf{q-k}}^{i\epsilon_n-i\omega_m}
\Gamma_{\mathbf{k},\mathbf{k}_2},
\end{eqnarray}
where the Matsubara Green's function is given as
\begin{eqnarray}
\mathcal{G}(\mathbf{k},i\epsilon_n) &=& \frac{1}{i\hbar \epsilon_n-\xi_{\mathbf{k}}+i\frac{\hbar}{2\tau}\mathrm{sgn}(\epsilon_n)}, \end{eqnarray}
the fermionic Matsubara frequency $\epsilon_n=(2n+1)\pi k_B T/\hbar$ with $n=0,\pm 1,\pm 2,...$, the bosonic Matsubara frequency $\omega_m=2\pi m k_B T/\hbar$ with $m=0,\pm 1,\pm 2,...$, and $\xi_\mathbf{k}=E_\mathbf{k}-E_F$ with $E_F$ the Fermi energy.
The bare Cooperon
\begin{eqnarray}
&& \Gamma_{\mathbf{k}_1,\mathbf{k}_2}^0 \equiv \langle U_{\mathbf{k}_1,\mathbf{k}_2}U_{-\mathbf{k}_1,-\mathbf{k}_2}\rangle\nonumber\\
&\approx &  \frac{\hbar(1-\eta_I)}{2\pi N_F\tau }  [\frac{1}{2} \sin\theta_1 \sin\theta_2+e^{i(\varphi_2-\varphi_1)}
\nonumber\\
&&+\cos\theta_1\cos\theta_2e^{i(\varphi_2-\varphi_1)}+\frac{1}{2} \sin\theta_1 \sin\theta_2 e^{i2(\varphi_2-\varphi_1)}],\nonumber\\
\end{eqnarray}
where $\eta_I\equiv \tau/\tau_I$ measures the relative strength of the intervalley scattering, and it can be found that
\begin{eqnarray}
&&\int_0^\infty \frac{dk k^2}{2\pi}
\mathcal{G}_\mathbf{k}^{i\epsilon_n}
\mathcal{G}_{\mathbf{q-k}}^{i\epsilon_n-i\omega_m}\nonumber\\
&\approx&\frac{2\pi^2 N_F \tau}{\hbar} \frac{1}{1+\omega_m \tau + i\tau \mathbf{v}_F\cdot \mathbf{q}}\nonumber\\
&\approx &
1-\omega_m \tau  - i\tau v_F q\cos\theta
-\tau^2 v_F^2 q^2 \cos^2\theta,
\end{eqnarray}
where $q^2=q_x^2+q_y^2+q_z^2$, which is essentially different from Ref. \onlinecite{Garate12prb}, where $q_z=0$ in a thin film with thickness $W\ll \ell_\phi$.
We can assume the form of the intravalley Cooperon
\begin{eqnarray}
\Gamma_{\mathbf{k}_1,\mathbf{k}_2}
&= & \frac{\hbar}{2\pi N_F\tau }  [c_1 \sin\theta_1 \sin\theta_2
+c_2e^{i(\varphi_2-\varphi_1)}\nonumber\\
&&+c_3\cos\theta_1\cos\theta_2e^{i(\varphi_2-\varphi_1)}\nonumber\\
&&+c_4 \sin\theta_1 \sin\theta_2 e^{i2(\varphi_2-\varphi_1)}\nonumber\\
&&+c_5\cos\theta_1e^{i(\varphi_2-\varphi_1)}+c_6\cos\theta_2e^{i(\varphi_2-\varphi_1)}].\nonumber\\
\end{eqnarray}
By putting it into the Bethe-Salpeter equation, we can find that
only the $c_2$ term is divergent as $\omega_m,q\rightarrow 0$ and
\begin{eqnarray}
\Gamma_{\mathbf{k}_1,\mathbf{k}_2}
&\approx  &  \frac{\hbar}{2\pi N_FD\tau^2 } \frac{2+ \eta_I}{2}
\frac{e^{i(\varphi_2-\varphi_1)}}{q^2+Q_0^2},
\end{eqnarray}
where
\begin{eqnarray}
Q_0^2 \equiv
\frac{(2+\eta_I)\eta_I}{2(1-\eta_I)}\frac{1}{\ell^2}.
\end{eqnarray}
As $\eta_I\rightarrow 0$,
\begin{eqnarray}
\Gamma_{\mathbf{k}_1,\mathbf{k}_2}
&= & \frac{\hbar}{2\pi N_F D\tau^2  }  \frac{1}{ q^2}e^{i(\varphi_2-\varphi_1)}.
\end{eqnarray}
In the bare Hikami box, $\mathbf{k}_1=\mathbf{k}$ and $\mathbf{k}_2=\mathbf{q}-\mathbf{k}\approx -\mathbf{k}$, then $\varphi_2=\pi+\varphi_1$, $e^{i(\varphi_2-\varphi_1)}=e^{i\pi}=-1$.
Similarly, in the dressed Hikami box, $e^{i(\varphi_2-\varphi_1)}$ becomes $e^{i(\varphi-\varphi_1)}$.

\subsection{Conductivity correction from intervalley Cooperons}\label{sec:qi-inter-cal}

\begin{figure}[tbph]
\centering \includegraphics[width=0.48\textwidth]{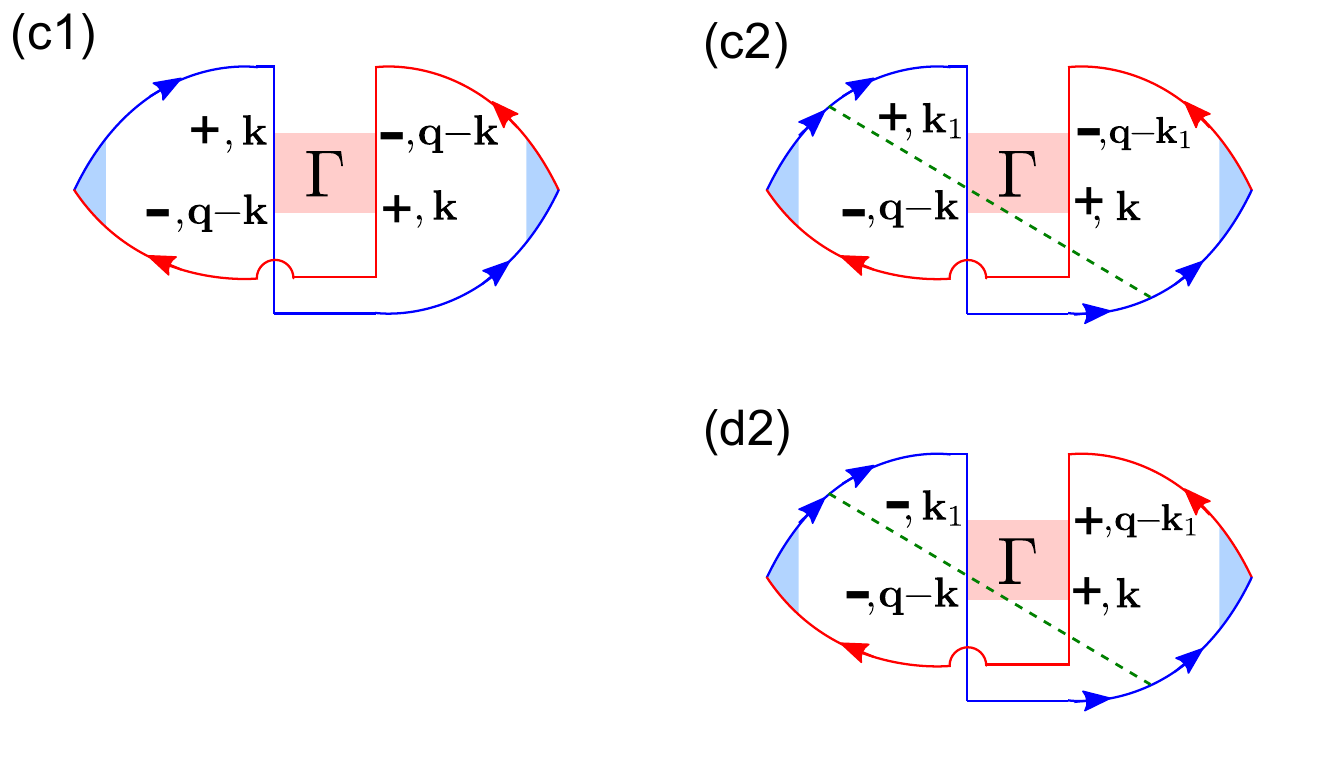}
\caption{The Feynman diagrams for the quantum interference correction to the conductivity that take into account the Cooperons from the intervalley scattering \cite{Lu13prl}. These diagrams give $\sigma^{qi}_I$.}
\label{fig:hikami-boxes-inter}
\end{figure}

The conductivity contribution from the intervalley Cooperons is given by (see Fig. \ref{fig:hikami-boxes-inter})
\begin{eqnarray}
\sigma^{qi}_I=2\times (\sigma_{c1}+2\times \sigma_{c2}+2\times \sigma_{d2}),
\end{eqnarray}
where
\begin{eqnarray}
\sigma_{c1}&=&\frac{e^2  \hbar}{2\pi}\sum_{\mathbf{q}}
\Gamma^{+-}_{-+}(\mathbf{k},-\mathbf{k})\nonumber\\
&&\times \sum_{\mathbf{k}}\widetilde{v}^x_{\mathbf{k},+}
G^R_{\mathbf{k},+}
G^R_{\mathbf{q}-\mathbf{k},-}
\widetilde{v}^x_{\mathbf{q}-\mathbf{k},-}
G^A_{\mathbf{q}-\mathbf{k},-}
G^A_{\mathbf{k},+},
\nonumber\\
\sigma_{c2}&=&\frac{e^2  \hbar}{2\pi}\sum_{\mathbf{q}}
\Gamma^{+-}_{-+}(\mathbf{k}_1,-\mathbf{k})\nonumber\\
&&\times \sum_{\mathbf{k}}\sum_{\mathbf{k}_1}
\widetilde{v}^x_{\mathbf{k},+}
G^R_{\mathbf{k},+}
G^R_{\mathbf{k}_1,+}
G^R_{\mathbf{q-k},-}
G^R_{\mathbf{q-k_1},-}\nonumber\\
&&\times \widetilde{v}^x_{\mathbf{q}-\mathbf{k}_1,-}
G^A_{\mathbf{q}-\mathbf{k}_1,-}
G^A_{\mathbf{k},+}
\langle U^{++}_{\mathbf{k},\mathbf{k}_1}U^{--}_{\mathbf{q-k},\mathbf{q-k_1}}\rangle , \nonumber\\
\sigma_{d2}&=&\frac{e^2  \hbar}{2\pi} \sum_{\mathbf{q}}\Gamma^{--}_{++}(\mathbf{k}_1,-\mathbf{k}) \nonumber\\ &&\times\sum_{\mathbf{k}}\sum_{\mathbf{k}_1}
\widetilde{v}^x_{\mathbf{k},t}
G^R_{\mathbf{k},+}G^R_{\mathbf{k}_1,-}
G^R_{\mathbf{q}-\mathbf{k},-}G^R_{\mathbf{q}-\mathbf{k}_1,+}
\nonumber\\
&&\times \widetilde{v}^x_{\mathbf{q}-\mathbf{k}_1,+}
G^A_{ \mathbf{q}-\mathbf{k}_1,+}
G^A_{\mathbf{k},+}
\langle U^{+-}_{\mathbf{k},\mathbf{k}_1}
U^{-+}_{\mathbf{q}-\mathbf{k},\mathbf{q}-\mathbf{k}_1}\rangle .
\end{eqnarray}

\begin{figure}[tbph]
\centering
\includegraphics[width=0.35\textwidth]{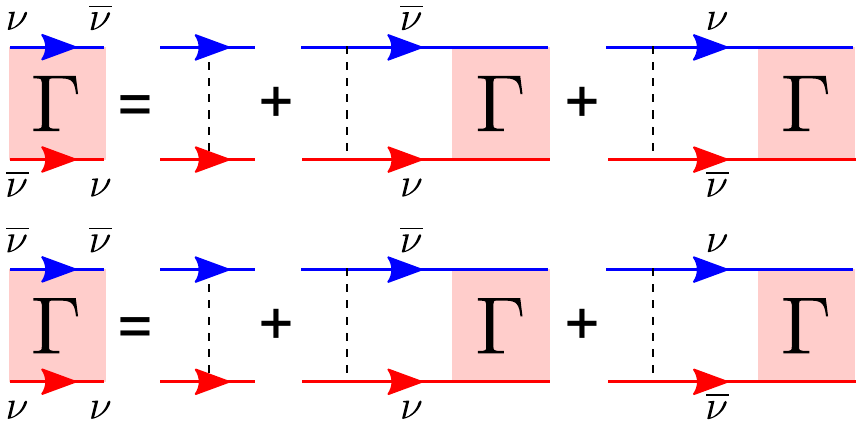}
\caption{The Feynman diagrams of the Bethe-Salpeter equations for the intervalley Cooperons. $\nu$ is the valley index, $\nu=\pm$, $\overline{\nu}=-$ if $\nu=+$.}
\label{fig:Cooperon-inter}
\end{figure}

In polar coordinates, the intervalley Cooperons can be found from the coupled Bethe-Salpeter equations (see Fig. \ref{fig:Cooperon-inter})
\begin{eqnarray}
&&\Gamma^{+-}_{-+}(\mathbf{k}_1,\mathbf{k}_2)\nonumber\\
&=&\gamma^{+-}_{-+}(\mathbf{k}_1,\mathbf{k}_2)
+\int_0^{2\pi}\frac{d\varphi}{2\pi}
\int_0^\pi \frac{d\theta\sin\theta}{2\pi}
\nonumber\\
&&\times\int_0^\infty \frac{dk k^2}{2\pi}\sum_{\nu=\pm}\gamma^{+\nu}_{-\overline{\nu}}(\mathbf{k}_1,\mathbf{k})
\mathcal{G}_\mathbf{k}^{i\epsilon_n}
\mathcal{G}_{\mathbf{q-k}}^{i\epsilon_n-i\omega_m}\Gamma^{\nu-}_{\overline{\nu}+}(\mathbf{k},\mathbf{k}_2),\nonumber\\
&&\Gamma^{--}_{++}(\mathbf{k}_1,\mathbf{k}_2)\nonumber\\
&=&\gamma^{--}_{++}(\mathbf{k}_1,\mathbf{k}_2)
+\int_0^{2\pi}\frac{d\varphi}{2\pi}
\int_0^\pi \frac{d\theta\sin\theta}{2\pi}
\nonumber\\
&&\times\int_0^\infty \frac{dk k^2}{2\pi}\sum_{\nu=\pm}
\gamma^{-\nu}_{+\overline{\nu}}(\mathbf{k}_1,\mathbf{k})
\mathcal{G}_\mathbf{k}^{i\epsilon_n}
\mathcal{G}_{\mathbf{q-k}}^{i\epsilon_n-i\omega_m}
\Gamma^{\nu-}_{\overline{\nu}+}(\mathbf{k},\mathbf{k}_2),\nonumber\\
\end{eqnarray}
where $\overline{\nu}=-$ if $\nu=+$ and
\begin{eqnarray}\label{g+--+}
&&\gamma^{\nu\overline{\nu}}_{\overline{\nu}\nu}
(\mathbf{k}_1,\mathbf{k}_2)\equiv \langle U^{\nu\overline{\nu}}_{\mathbf{k}_1,\mathbf{k}_2}
U^{\overline{\nu}\nu}_{-\mathbf{k}_1,-\mathbf{k}_2}\rangle\nonumber\\
&=& \frac{\hbar \eta_I}{2\pi N_F \tau}
[\frac{1}{2}(1+\nu\cos\theta_1)(1
+\overline{\nu}\cos\theta_2)\nonumber\\
&&-\sin\theta_1\sin\theta_2  e^{i(\varphi_2-\varphi_1)}\nonumber\\
&&+\frac{1}{2}(1+\overline{\nu}\cos\theta_1)(1
+\nu\cos\theta_2)e^{i2(\varphi_2-\varphi_1)}],\nonumber\\
&&\gamma^{\nu\nu}_{\overline{\nu}\overline{\nu}}(\mathbf{k}_1,\mathbf{k}_2) \equiv \langle U^{\nu\nu}_{\mathbf{k}_1,\mathbf{k}_2}U^{\overline{\nu}\overline{\nu}}_{-\mathbf{k}_1,-\mathbf{k}_2}\rangle\nonumber\\
&=& \frac{\hbar \eta_*}{2\pi N_F \tau}
[\frac{1}{2}(1+\nu\cos\theta_1)(1+\nu\cos\theta_2) \nonumber\\
&&+\sin\theta_1\sin\theta_2  e^{i(\varphi_2-\varphi_1)}\nonumber\\ &&+\frac{1}{2}(1
+\overline{\nu} \cos\theta_1)(
1+\overline{\nu}\cos\theta_2)e^{i2(\varphi_2-\varphi_1)}],
\end{eqnarray}
where $\nu=\pm $, $\eta_I\equiv \tau/\tau_I$, and $\eta_*\equiv \tau/\tau_*$.
We assume
\begin{eqnarray}\label{Gamma-trial+--+}
\Gamma^{+-}_{-+}(\mathbf{k}_1,\mathbf{k}_2)&=& \frac{\hbar  }{2\pi N_F \tau}
[a_1(1+\cos\theta_1)(1-\cos\theta_2)\nonumber\\
&&-a_2\sin\theta_1\sin\theta_2  e^{i(\varphi_2-\varphi_1)}\nonumber\\
&&+a_3 (1-\cos\theta_1)(1+\cos\theta_2)e^{i2(\varphi_2-\varphi_1)}] ,\nonumber\\
\Gamma^{--}_{++}(\mathbf{k}_1,\mathbf{k}_2)&=&
\frac{\hbar  }{2\pi N_F \tau}
[b_1(1-\cos\theta_1)(1-\cos\theta_2)\nonumber\\
&&+b_2\sin\theta_1\sin\theta_2  e^{i(\varphi_2-\varphi_1)}\nonumber\\
&&+b_3(1+\cos\theta_1)(1+\cos\theta_2)e^{i2(\varphi_2-\varphi_1)}],\nonumber\\
\end{eqnarray}
and put them into the Bethe-Salpeter equations, we arrive at
\begin{eqnarray}
a_1=a_3
&=& \frac{1}{2\ell^2}\frac{\chi_1^a}{Q_1^2+q^2},\ \
a_2 = \frac{1}{\ell^2} \frac{\chi_2^a}{Q_2^2+q^2},
\nonumber\\
b_1=b_3
&=& \frac{1}{2\ell^2}\frac{\chi_1^b}{Q_1^2+q^2},\ \
b_2 =
\frac{1}{\ell^2}\frac{\chi_2^b}{Q_2^2+q^2},
\end{eqnarray}
where $q^2=q_x^2+q_y^2+q_z^2$ and other quantities are defined in Eq. (\ref{def-QXW}).

\subsection{Conductivity and magnetoconductivity from quantum interference}\label{sec:qi-c-mc}

The total conductivity from the quantum interference has two parts
\begin{eqnarray}
\sigma^{qi} &=& 2\times \sigma^{qi}_0+\sigma^{qi}_I.
\end{eqnarray}
For a single valley,
\begin{eqnarray}\label{sigma-qi-intra-1}
\sigma^{qi}_{0}
&\approx & \frac{e^2}{h}\frac{2+ \eta_I}{(1+\eta_I)^2}
\sum_{\mathbf{q}}
\frac{1}{Q_0^2 +q^2},
\end{eqnarray}
where
\begin{eqnarray}
Q_0^2 \equiv
\frac{(2+\eta_I)\eta_I}{2(1-\eta_I)}\frac{1}{\ell^2}
\end{eqnarray}
and $\eta_I=\tau/\tau_I$ measures the weight of the intervalley scattering in the total scattering. The total scattering time $\tau$ is defined as $1/\tau=1/\tau_0+1/\tau_I$, $\tau_0$ and $\tau_I$ are the intravalley and intervalley scattering times, respectively.

The intervalley part
\begin{eqnarray}
\sigma^{qi}_I=2\times (\sigma_{c1}+2\times \sigma_{c2}+2\times \sigma_{d2}),
\end{eqnarray}
where
\begin{eqnarray}
\sigma_{c1}^x
 &=& -\frac{e^2}{h}\frac{1}{(1+\eta_I)^2}
\sum_{\mathbf{q}}(\frac{18}{5}\frac{\chi_1^a}{Q_1^2+q^2}
+\frac{12}{5}\frac{\chi_2^a}{Q_2^2+q^2}),\nonumber\\
\sigma_{c2}^x
&=&\frac{e^2}{h}\frac{1}{(1+\eta_I)^2}
\sum_{\mathbf{q}}
\frac{\eta_*}{2}(\frac{\chi_1^a}{Q_1^2+q^2}
+\frac{\chi_2^a}{Q_2^2+q^2} ),\nonumber\\
\sigma_{d2}^x &=&-\frac{e^2  }{h}\frac{1}{(1+\eta_I)^2}
\sum_{\mathbf{q}}
\frac{\eta_I}{ 2}
(\frac{\chi_1^b}{Q_1^2+q^2}+ \frac{\chi_2^b}{Q_2^2+q^2}),
\end{eqnarray}
and $q^2=q_x^2+q_y^2+q_z^2$,
\begin{eqnarray}\label{def-QXW}
Q_1^2&=&\frac{(1-\frac{2}{3}\eta_*)^2
-(\frac{2}{3}\eta_I)^2}{\varpi_1\ell^2},\
Q_2^2=
\frac{ (1- \frac{2}{3}\eta_*)^2
-(\frac{2}{3}\eta_I)^2}{\varpi_2\ell^2},
\nonumber\\
\chi_1^a
&=&\eta_I/\varpi_1,\ \
\chi_1^b
= [\eta_* +
2(\eta_I^2-\eta_*^2)/3]/
\varpi_2,\nonumber\\
\chi_2^a &=&
\eta_I/\varpi_2,\ \
\chi_2^b =
[\eta_*
 +2(\eta_I^2-\eta_*^2)/3]/
\varpi_2,\nonumber\\
\varpi_1 &=&   \frac{16}{15}\eta_*
+\frac{22}{45}(\eta_I^2-\eta_*^2),\
\varpi_2= \frac{8}{15}\eta_*
+\frac{16}{45}(\eta_I^2-\eta
_*^2),\nonumber\\
\eta_I&=&\tau/\tau_I,\ \ \eta_*=\tau/\tau_*
.
\end{eqnarray}
One can check that $\sigma^{qi}_I$ vanishes when $\eta_I=0$.

To have the temperature dependence of the conductivity,
one just replaces all the $ \sum_\mathbf{q} \frac{1}{Q_i^2+q^2}$
in $\sigma^{qi}$ by
\begin{eqnarray}\label{q-integral}
\frac{1}{2\pi^2}\int_{1/\ell_\phi}^{1/\ell}
q^2 dq.
\end{eqnarray}
The temperature dependence is contained in the phase coherence length $\ell_\phi = C /T^{p/2}$, $C$ is a constant. In three dimensions, $p=3/2$ ($p=3$) if the electron-electron (electron-phonon) interaction is the decoherence mechanism \cite{Lee85rmp}.

To calculate the magnetoconductivity, one just replaces all the
 $ \sum_\mathbf{q} \frac{1}{Q_i^2+q^2}$
in $\sigma^{qi}$ by
\begin{eqnarray}\label{MC-cal}
\Psi_3(B,Q_i)
&=&   \int_{0}^{1/\ell}
\frac{dx}{(2\pi)^2}
\left[ \psi\left(\frac{\ell_B^2}{\ell^2}+\ell_B^2(Q_i^2+x^2)+\frac{1}{2}\right) \right.\nonumber\\ &&\left.-\psi\left(\frac{\ell_B^2}{\ell_\phi^2}+\ell_B^2(Q_i^2+x^2)
+\frac{1}{2}\right)  \right],
\end{eqnarray}
where the magnetic length $\ell_B\equiv \sqrt{\hbar/ 4eB }$, the magnetic field $B$ is along arbitrary directions.
The magnetoconductivity is defined as
\begin{eqnarray}
\delta\sigma^{qi}(B)\equiv \sigma^{qi}(B)-\sigma^{qi}(0).
\end{eqnarray}

\subsection{Conductivity and magnetoconductivity of a single valley of Weyl fermions}

For a single valley in absence of intervalley scattering, $\sigma^{qi}$ reduces to $\sigma^{qi}_0$ in Eq. (\ref{sigma-qi-intra-1}) with $\eta_I=0$,
\begin{eqnarray}\label{sigma-qi-0-1}
\sigma^{qi}= \frac{e^2}{h}2
\sum_{\mathbf{q}}
\frac{1}{q^2}.
\end{eqnarray}
Replace the summation by the integral in Eq. (\ref{q-integral}),
\begin{eqnarray}
\sigma^{qi}=\frac{e^2}{h}2
\frac{1}{2\pi^2}\int_{1/\ell_\phi}^{1/\ell}
dq \frac{1}{q^2}q^2
=\frac{e^2}{h}\frac{1}{\pi^2}(\frac{1}{\ell}-\frac{1}{\ell_\phi}).
\end{eqnarray}
This is of the same magnitude of Eq. (2.25a) of Ref. \onlinecite{Lee85rmp} but differs by a minus sign (note that spin degeneracy 2 is included in Ref. \onlinecite{Lee85rmp}).

Replace the summation in Eq. (\ref{sigma-qi-0-1}) by Eq. (\ref{MC-cal}), $\sigma^{qi}(B)$ of a single valley is found as
\begin{eqnarray}
\sigma^{qi}(B)= \frac{e^2}{h}2
\Psi_3(B),
\end{eqnarray}
and the magnetoconductivity is
\begin{eqnarray}
\delta\sigma^{qi}(B)&\equiv &\sigma^{qi}(B)-\sigma^{qi}(0),\nonumber\\
\end{eqnarray}
where
\begin{eqnarray}
\sigma^{qi}(B)
&=& \frac{2e^2}{h}  \int_{0}^{1/\ell}
\frac{dx}{(2\pi)^2}
\left[ \psi\left(\frac{\ell_B^2}{\ell^2}+\ell_B^2x^2 +\frac{1}{2}\right) \right.\nonumber\\ &&\left.-\psi\left(\frac{\ell_B^2}{\ell_\phi^2}+\ell_B^2 x^2
+\frac{1}{2}\right)  \right],
\end{eqnarray}
where the magnetic length $\ell_B\equiv \sqrt{\hbar/ 4eB }$, the magnetic field $B$ is along arbitrary directions, $\ell$ is the mean free path, and $\ell_\phi$ is the phase coherence length.

\section{Conductivity correction from interaction}\label{sec:ee-cal}

\begin{figure}[tbph]
\centering
\includegraphics[width=0.48\textwidth]{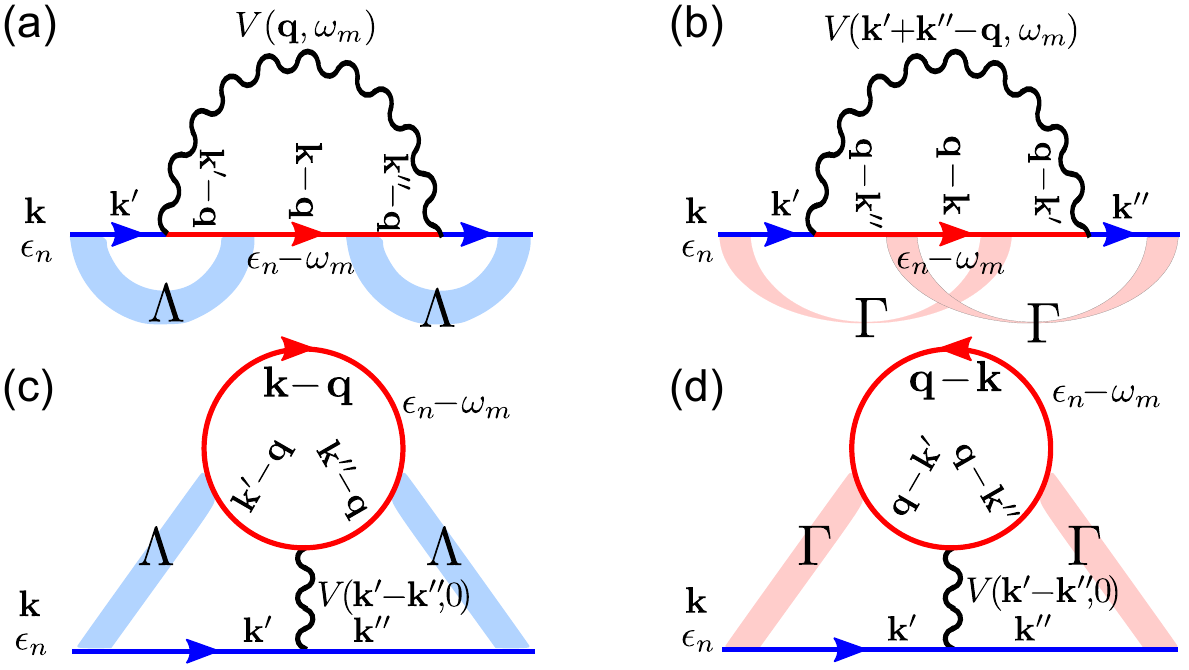}
\caption{The Feynman diagrams for the self-energies arising from the interplay of electron-electron interaction and disorder scattering \cite{Altshuler80prl,Fukuyama80jpsj}.}
\label{fig:one-loop}
\end{figure}

The leading-order of the self-energy from the interplay of interaction and disorder is the Fock (exchange) diagram dressed by Diffusons [see Fig. \ref{fig:one-loop}(a)],
\begin{eqnarray}\label{Sigma-DF-def}
\Sigma^{F}_D(\mathbf{k},i\epsilon_n) &=&
-\frac{1}{\beta}\sum_{\omega_m}\sum_{\mathbf{q}}\mathcal{G}_{\mathbf{k}-\mathbf{q}}^{i\epsilon_n-i\omega_m}
\sum_{\mathbf{k}'}\sum_{\mathbf{k}''}
V(\mathbf{q},i\omega_m)\nonumber\\
&&\times
\Lambda^{\mathbf{k}\rightarrow \mathbf{k}',i\epsilon_n}
_{\mathbf{k}-\mathbf{q}\leftarrow \mathbf{k}'-\mathbf{q},i\epsilon_n-i\omega_m}
\Lambda^{\mathbf{k}''\rightarrow \mathbf{k},i\epsilon_n}
_{\mathbf{k}''-\mathbf{q}\leftarrow \mathbf{k}-\mathbf{q},i\epsilon_n-i\omega_m}\nonumber\\
&&\times \mathcal{G}_{\mathbf{k}'}^{i\epsilon_n}
\mathcal{G}_{\mathbf{k}'-\mathbf{q}}^{i\epsilon_n-i\omega_m}
\mathcal{G}_{\mathbf{k}''-\mathbf{q}}^{i\epsilon_n-i\omega_m}
\mathcal{G}_{\mathbf{k}''}^{i\epsilon_n} ,
\end{eqnarray}
where $1/\beta=k_BT$, later we will show how to calculate the Diffuson $\Lambda$ and interaction $V$. We find that (Sec. \ref{sec:Diffuson-cal})
\begin{eqnarray}
&&\Lambda^{\mathbf{k}\rightarrow \mathbf{k}',i\epsilon_n}
_{\mathbf{k}-\mathbf{q}\leftarrow \mathbf{k}'-\mathbf{q},i\epsilon_n-i\omega_m}
=\Lambda^{\mathbf{k}''\rightarrow \mathbf{k},i\epsilon_n}
_{\mathbf{k}''-\mathbf{q}\leftarrow \mathbf{k}-\mathbf{q},i\epsilon_n-i\omega_m}\nonumber\\
&=&\Lambda_{\mathbf{k}_1,\mathbf{k}_2}
\approx \frac{\hbar}{2\pi N_F\tau^2}
\frac{1}{\omega_m+Dq^2},
\end{eqnarray}
and
\begin{eqnarray}
\sum_{\mathbf{k}'}
\mathcal{G}_{\mathbf{k}'}^{i\epsilon_n}
\mathcal{G}_{\mathbf{k}'-\mathbf{q}}^{i\epsilon_n-i\omega_m}
\approx \frac{2\pi N_F \tau}{\hbar} \theta[\epsilon_n(\omega_m-\epsilon_n)].
\end{eqnarray}
Then
\begin{eqnarray}
&&\Sigma^{F}_D(\mathbf{k},i\epsilon_n) \nonumber\\
&\approx &
-\frac{1}{\beta}\sum_{\omega_m}
\mathcal{G}_{\mathbf{k}}^{i\epsilon_n-i\omega_m}
\frac{1}{\tau^2}
\sum_{\mathbf{q}}
\frac{V_{\mathbf{q}}^{i\omega_m}\theta[\epsilon_n(\omega_m-\epsilon_n)]}{(\omega_m+Dq^2)^2}.\nonumber\\
\end{eqnarray}
It is of the same form as Eq. (3.16) in Ref. \onlinecite{Altshuler-book}. Therefore, the leading-order self-energy and its associated contribution to the conductivity has the same form as that for the conventional electron with dispersion $p^2/2m$. The difference is that $D$ and $N_F$ need to be changed to those for the Weyl fermions.

\subsection{Diffuson}\label{sec:Diffuson-cal}

\begin{figure}[tbph]
\centering
\includegraphics[width=0.3\textwidth]{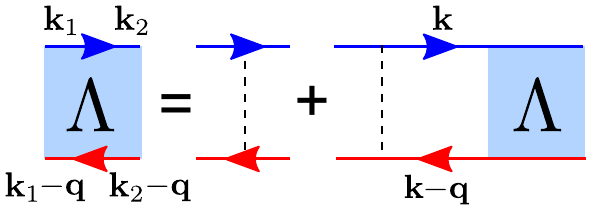}
\caption{The Feynman diagrams of the Bethe-Salpeter equations for the Diffuson.}
\label{fig:Diffuson}
\end{figure}

In polar coordinates, the Diffuson can be found from the Bethe-Salpeter equation (see Fig. \ref{fig:Diffuson})
\begin{eqnarray}
\Lambda_{\mathbf{k}_1,\mathbf{k}_2}
&=&\Lambda_{\mathbf{k}_1,\mathbf{k}_2}^0
+\int_0^{2\pi}\frac{d\varphi}{2\pi}
\int_0^\pi \frac{d\theta\sin\theta}{2\pi}\int_0^\infty \frac{dk k^2}{2\pi}\nonumber\\
&&\times \Lambda_{\mathbf{k}_1,\mathbf{k}}^0 \mathcal{G}_\mathbf{k}^{i\epsilon_n}\mathcal{G}_{\mathbf{k-q}}^{i\epsilon_n-i\omega_m}\Lambda_{\mathbf{k},\mathbf{k}_2},
\end{eqnarray}
where the bare Diffuson can be found
\begin{eqnarray}
&&\Lambda^0_{\mathbf{k}_1,\mathbf{k}_2}\equiv \langle U_{\mathbf{k}_1,\mathbf{k}_2}U_{\mathbf{k}_2,\mathbf{k}_1}\rangle\nonumber\\
&\approx &  \frac{\hbar}{2\pi N_F\tau}[1+(1-2\eta_I)\nonumber\\
&&\times(\cos\theta_1\cos\theta_2+\sin\theta_1\sin\theta_2\cos(\varphi_1-\varphi_2))],
\end{eqnarray}
and it can be found that
\begin{eqnarray}
\int_0^\infty \frac{dk k^2}{2\pi}
\mathcal{G}_\mathbf{k}^{i\epsilon_n}
\mathcal{G}_{\mathbf{k-q}}^{i\epsilon_n-i\omega_m}=\frac{2\pi^2 N_F \tau}{\hbar} \frac{1}{1+\omega_m \tau + i\tau \mathbf{v}_F\cdot \mathbf{q}}.\nonumber\\
\end{eqnarray}
For convenience, the $z$-axis of $\mathbf{k}$ can be chosen to be along the direction of $\mathbf{q}$, then $\mathbf{v}_F\cdot \mathbf{q}=v_F q \cos\theta$ and
\begin{eqnarray}
&&\frac{1}{1+\omega_m\tau +i \tau \mathbf{v}_F\cdot \mathbf{q}}\nonumber\\
&\approx &
1-\omega_m \tau  - i\tau v_F q\cos\theta
-\tau^2 v_F^2 q^2 \cos^2\theta.
\end{eqnarray}
We assume the form of the full Diffuson to be
\begin{eqnarray}
\Lambda_{\mathbf{k}_1,\mathbf{k}_2}
&\approx & \frac{\hbar}{2\pi N_F\tau}[d_1+d_2\cos\theta_1+d_3\cos\theta_2\nonumber\\
&&+d_4\cos\theta_1\cos\theta_2+d_5\sin\theta_1\sin\theta_2\cos(\varphi_1-\varphi_2)],\nonumber\\
\end{eqnarray}
with the coefficients $d_i$ to be determined. By putting it into the Bethe-Salpeter equation, we find that only the $d_1$ term is divergent as $q,\omega_m\rightarrow 0$ and
\begin{eqnarray}
\Lambda_{\mathbf{k}_1,\mathbf{k}_2}
&\approx & \frac{\hbar}{2\pi N_F\tau^2}
\frac{1}{\omega_m+Dq^2},
\end{eqnarray}
where the diffusion coefficient
\begin{eqnarray}
D=\frac{1}{3}v_F^2\tau \eta_v.
\end{eqnarray}
It is worth noting that here the expression of $D$ derived from the Diffuson coincides with that in the semiclassical conductivity (Sec. \ref{sec:sc-cal}). The above calculation does not distinguish inter- and intra-valley scattering $\Lambda=\Lambda^{++}_{++}+\Lambda^{+-}_{+-}$.

\subsection{Interaction and random phase approximation}

\begin{figure}[tbph]
\centering
\includegraphics[width=0.3\textwidth]{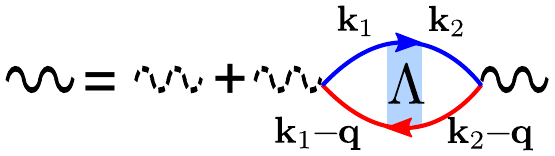}
\caption{The Feynman diagram for the interaction (dashed wavy lines for bare interaction), which is renormalized (solid wavy lines) under random phase approximation. Different from in a clean system, the density function is dressed by the Diffuson ($\Lambda$).}
\label{fig:RPA}
\end{figure}

\begin{figure}[tbph]
\centering
\includegraphics[width=0.45\textwidth]{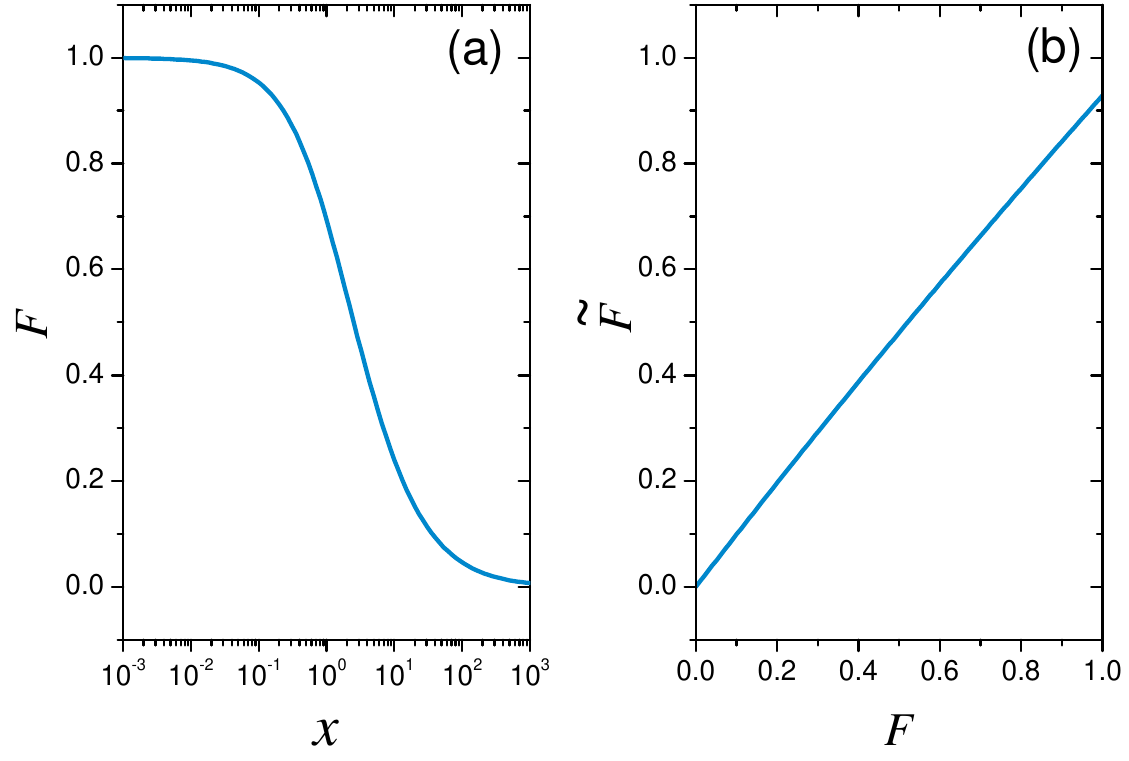}
\caption{(a) The screening factor $F$ as a function of $x=8\pi^2 v_F\hbar \varepsilon /e^2$ for Weyl fermions. $v_F$ is the Fermi velocity. $\hbar $ is the reduced Planck's constant. $\varepsilon$ is the dielectric constant. $-e$ is the electron charge. (b) In three dimensions, $\widetilde{F}$ as a function of $F$. $\tilde{F}$ is the renormalized screening factor after including the second-order diagrams and correction of the disorder by the interaction.}
\label{fig:F}
\end{figure}

After the Fourier transformation, the Hamiltonian of the interaction becomes
\begin{eqnarray}
V = \sum_{\mathbf{k},\mathbf{k}',\mathbf{q}}
\frac{V_{\mathbf{q}}}{2}(\phi_{\mathbf{k}}^\dag \cdot \phi_{\mathbf{k}+\mathbf{q}}) (\phi_{\mathbf{k}'}^\dag\cdot
\phi_{\mathbf{k}'-\mathbf{q}})
c^{\dag}_{\mathbf{k}'}c^{\dag}_{\mathbf{k}} c_{\mathbf{k}+\mathbf{q}} c_{\mathbf{k}'-\mathbf{q}},\nonumber\\
\end{eqnarray}
where $\phi$'s are the spinor wave functions, $c$'s are corresponding operators, and
\begin{eqnarray}
V_{\mathbf{q}}= \int \mathrm{d}^3 \mathbf{r} \frac{e^2}{4\pi \varepsilon r} e^{-
        \mathrm{i} \mathbf{q} \cdot \mathbf{r}}
          =   \frac{e^2}{\varepsilon q^2 }
\end{eqnarray}
with $\varepsilon$ the dielectric constant.
Because $v_{q}$ diverges as $q\rightarrow 0$, then the spinor wave function part vanishes for a single band in the interaction potential
\begin{eqnarray}
V_0(q)=v_{\mathbf{q}}(\phi_{\mathbf{k}}^\dag \cdot \phi_{\mathbf{k}+\mathbf{q}}) (\phi_{\mathbf{k}'}^\dag\cdot
\phi_{\mathbf{k}'-\mathbf{q}})
\approx V_{\mathbf{q}}.
\end{eqnarray}
The long-range (bare) interaction is renormalized under the random phase approximation (see Fig. \ref{fig:RPA})
\begin{eqnarray}
V(\mathbf{q},\omega_m) &=& \frac{V_0(q)}{1+V_0(q)\Pi(q,\omega_m)},
\end{eqnarray}
where different from in a clean system, the density response function in a disordered system is dressed by the Diffuson and takes the form
\begin{eqnarray}
\Pi(\mathbf{q},\omega_m) &=& N_F\frac{Dq^2}{\omega_m+Dq^2}.
\end{eqnarray}
Then
\begin{eqnarray}
V(\mathbf{q},i\omega_m) &=& \frac{e^2}{\varepsilon q^2 +e^2N_F\frac{Dq^2}{\omega_m+Dq^2} }.
\end{eqnarray}
In the limit that $\omega_m,\ q\rightarrow 0$, the dynamically-screened interaction becomes
\begin{eqnarray}
V(\mathbf{q},i\omega_m) \approx \frac{1}{N_F}\frac{\omega_m+Dq^2}{Dq^2}.
\end{eqnarray}
This renormalized interaction is the one that is used in calculating the self-energy induced by the interplay of interaction and disorder.

\subsection{Screening factor}\label{sec:F-cal}

The contribution from other three one-loop interaction diagrams [see Fig. \ref{fig:one-loop} (b)-(d)] is proportional to the screening factor $F$, which is defined as
\begin{eqnarray}
F \equiv \frac{\langle V(\mathbf{k}-\mathbf{k}')\rangle_{k_F}}{V(0)},
\end{eqnarray}
where $\langle ...\rangle_{k_F}$ means the average of the interaction $V(\mathbf{k}-\mathbf{k}')$ over momenta $\mathbf{k}$ and $\mathbf{k}'$ on the Fermi surface.

In three dimensions \cite{Altshuler-book},
\begin{eqnarray}
F &=& \frac{\ln [(1+(2k_F\xi)^2]}{(2k_F\xi)^2},
\end{eqnarray}
where $k_F$ the Fermi wave vector and $\xi$ is the screening length of the interaction,
\begin{eqnarray}
\xi^2 = \frac{\varepsilon}{e^2N_F}.
\end{eqnarray}
Using the density of states for the Weyl fermions
$N_F= E_F^2/2\pi^2(v_F\hbar)^3$ and $E_F=v_F\hbar k_F$, we define
\begin{eqnarray}
x\equiv (2k_F\xi)^2= \frac{8\pi^2v_F\hbar\varepsilon}{e^2}.
\end{eqnarray}
Fig. \ref{fig:F} shows $F$ as a function of $x$. By definition, $F\in[0,1]$.

The screening factor $F$ will be renormalized after including the second-order diagrams and correction of the disorder by the interaction. The renormalized screening factor in three dimensions is \cite{Altshuler-book}
\begin{eqnarray}
\widetilde{F} = -\frac{32}{3} [1+\frac{3F}{4}-(1+\frac{F}{2})^{3/2}]F,
\end{eqnarray}
which is shown in Fig. \ref{fig:F} as a function of $F$. One can see that $\widetilde{F}\approx F$ as $F\rightarrow 0$ and $\widetilde{F}\approx 0.93 F$ as $F\rightarrow 1$.

\section{Classical magnetoconductivity from Lorentz force }\label{sec:sigma-c-cal}

The classical negative magnetoconductivity as a result of the cyclotron motion driven by Lorentz force in a perpendicular magnetic field $B$ can be found as \cite{Datta1997}
\begin{eqnarray}
\delta\sigma^C(B) &=& -\sigma^{sc}(\mu B)^2,
\end{eqnarray}
where $\sigma^{sc}$ is given by Eq. (\ref{Drude-L-kF}) and $\mu$ is given by Eq. (\ref{mobility}). We arrive at
\begin{eqnarray}
\delta\sigma^C(B) &=& -\frac{e^2}{h}\frac{\ell^3\sqrt{3}\eta_v^{3/2}}{16\pi\ell_B^4}\nonumber\\
&=&-\frac{e^2}{h}\frac{9 }{32\sqrt{2}\pi  (1+\eta_I)^{3/2}}\frac{\ell^3}{\ell_B^4},
\end{eqnarray}
where $\ell_{B}\equiv \sqrt{\hbar/4eB }$.

The mobility of one valley of Weyl fermion is found as
\begin{eqnarray}\label{mobility}
\mu &=& \frac{ev_F \tau \eta_v}{\hbar k_F }=\frac{e\ell\sqrt{3 \eta_v}}{\hbar k_F}
\approx\frac{e\ell}{\hbar k_F},
\end{eqnarray}
where
the mean free path $\ell\equiv \sqrt{D\tau}=v_F\tau \sqrt{\eta_v/3}\approx v_F \tau $.

The relation between the mobility and mean free path is approximated as
\begin{eqnarray}
\ell \approx \frac{\mu\hbar k_F}{e} \approx 66  \mu k_F,
\end{eqnarray}
where $\ell$ in nm, $\mu$ is in cm$^2$/(V$\cdot$s), and $k_F$ in $\AA^{-1}$.
For $k_F=0.03$ and $\mu=10^4$, the mean free path is about $20\mu$m.

\section{conclusions}

In this work, we study the quantum transport properties of a two-valley Weyl semimetal.
We employ the Feynman diagram techniques to calculate the conductivity
in the presence of disorder and interaction.
We derive three dominant parts of the conductivity (see Fig. \ref{fig:diagram}), including the semiclassical (Drude) conductivity, the correction from the quantum interference [weak (anti-)localization]
, and the correction
from the interplay of electron-electron interaction and disorder scattering (Altshuler-Aronov effect).

The quantum interference gives the main contribution to the magnetoconductivity. For a single valley of Weyl fermions, the low-temperature magnetoconductivity is proportional to $-\sqrt{B}$, where $B$ is the magnetic field applied along arbitrary directions [see Fig. \ref{fig:MC} (a)]. This $-\sqrt{B}$ magnetoconductivity is from the weak antilocalization of Weyl fermions in the presence of weak inter-valley scattering. Near zero field, the $-\sqrt{B}$ magnetoconductivity always overwhelms the positive $B^2$ magnetoconductivity from the chiral anomaly, giving another transport signature of Weyl semimetals. Strong inter-valley scattering and correlation can lead to a crossover from the weak antilocalization to weak localization. During the crossover, the $-\sqrt{B}$ magnetoconductivity turns to $\sqrt{B}$
in the limit of strong inter-valley scattering and correlation [see Fig. \ref{fig:MC} (c)]. By including the contributions from the weak antilocalization, Berry curvature correction, and Lorentz force (Tab. \ref{tab:MC}), we compare the calculated magnetoconductivity with a recent experiment (see Fig. \ref{fig:Kim}).

Both the quantum interference and interaction contribute to the temperature dependence of the conductivity.
For a single valley of Weyl fermions, the weak antilocalization from the quantum interference gives a conductivity proportional to $-T^{p/2}$ , where $T$ is the temperature and the parameter $p$ is positive and depends on decoherence mechanisms. This conductivity thus always increases with decreasing temperature, giving another signature of the weak antilocalization. In contrast, the interaction gives a conductivity that decreases with decreasing temperature, following a $\sqrt{T}$ dependence. Therefore, we expect a competition in the temperature dependence of the conductivity (see Fig. \ref{fig:sigma-T}). Because $p$ is usually greater than $1$, the interaction always dominates below a critical temperature, leading to a tendency to localization in the temperature-dependent conductivity.

We also present a systematic comparison of the transport properties for a single valley of Weyl fermions, 2D massless Dirac fermions, and 3D conventional electrons (Table \ref{tab:2D-3D}).

\begin{acknowledgements}
This work was supported by Research Grants Council, University Grants
Committee, Hong Kong, under Grant No. 17303714.
\end{acknowledgements}


%

\end{document}